\documentclass[12pt]{article}
\usepackage {epsfig}

\def\hpi{{h_\pi^1}}
\def\hrhoz{{h_\rho^0}}
\def\hrhoo{{h_\rho^1}}

\def\hrhoop{{h_\rho^{1\prime}}}
\def\homz{{h_\omega^0}}
\def\homo{{h_\omega^1}}

\newcommand{\gsim}{~{}_{\textstyle\sim}^{\textstyle >}~}
\newcommand{\lsim}{~{}_{\textstyle\sim}^{\textstyle <}~}

\newcommand{\be}{\begin{equation}}
\newcommand{\ee}{\end{equation}}
\newcommand{\bea}{\begin{eqnarray}}
\newcommand{\eea}{\end{eqnarray}}

\usepackage[sort&compress]{natbib}
\bibpunct{(}{)}{,}{n}{}{,}

\title{Hadronic Parity Violation:  a New View through the Looking Glass}

\vspace*{0.5cm}
\author{Michael J. Ramsey-Musolf\\
Kellogg Radiation Laboratory,\\ California Institute of
Technology,\\ Pasadena, CA 91125 USA\\
~~\\
and\\
~~\\
Shelley A. Page\\
Dept. of Physics \& Astronomy, \\
Univ. of Manitoba,\\
Winnipeg, MB R3T 2N2 Canada}

\makeatletter
\renewcommand\@biblabel[1]{#1.}
\makeatother

\input epsf.tex    %<-If you need EPS figures to be
\begin{document}

%\input psfig.sty

%\jname{Annu. Rev. Nucl. Part. Sci.} \jyear{2005} \jvol{55}
%\ARinfo{1056-8700/97/0610-00}

\maketitle

\begin{abstract}
Studies of the strangeness changing hadronic weak interaction have produced a number of puzzles that have so far evaded a complete explanation
within the Standard Model. Their origin may lie either in dynamics peculiar to weak interactions involving strange quarks or in more general
aspects of the interplay between strong and weak interactions. In principle, studies of the strangeness conserving hadronic weak interaction
using parity violating hadronic and nuclear observables provide a complementary window on this question. However, progress in this direction has
been hampered by the lack of a suitable theoretical framework for interpreting hadronic parity violation measurements in a model-independent
way. Recent work involving effective field theory ideas has led to the formulation of such a framework  while motivating the development of a
number of new hadronic parity violation experiments in few-body systems. In this article, we review these recent developments and discuss the
prospects and opportunities for further experimental and theoretical progress.

\end{abstract}

\pagebreak

\tableofcontents
\newpage

\section{Introduction} Explaining the weak interactions
of quarks in terms of the dynamics of the Standard Model (SM) has been an area of vigorous research in nuclear and particle physics for several
decades. Experimentally, the hadronic weak interaction (HWI) is probed by observing non-leptonic, flavor changing decays of mesons and baryons
and by measuring observables that conserve flavor but violate the parity symmetry of the strong and electromagnetic interactions. Theoretically,
the problem has been a particularly challenging one, requiring the computation of low-energy weak matrix elements of the HWI in strongly
interacting systems. Although the structure of the weak quark-quark interaction in the SM has been well established for some time, its
manifestation in strongly interacting systems remains only partially understood. The stumbling block has been the non-perturbative nature of
quantum chromodynamics (QCD) at low energies. In contending with it, theorists have resorted to a variety of approximation schemes to obtain
physically reasonable estimates of HWI observables. Ultimately, however,  arriving at definitive, SM predictions requires that one treat  the
non-perturbative QCD dynamics in a rigorous way.

In the case of the flavor changing decays of mesons, use of effective field theory (EFT) techniques -- chiral perturbation theory ($\chi$PT),
heavy quark effective theory (HQET), and recently, soft collinear effective theory (SCET) -- have led to enormous progress. In each instance,
the presence of distinct physical scales at play in the processes of interest allows one to carry out a systematic expansion of the effective
Lagrangian in powers of scale ratios while incorporating the symmetries of QCD into the structure of the operators. The operator coefficients
that encode the non-perturbative QCD dynamics are obtained from measurement, and the structure of the EFT is then used to translate this
information into predictions for other observables. Moreover, a meaningful confrontation of experiment with QCD theory can be made, as
computations of the operator coefficients can in principle be performed on the lattice.

The situation involving the HWI of baryons is far less satisfactory, and decades of experimental and theoretical work have left us with a number
of unresolved puzzles. In the case of hyperon non-leptonic decays, for example, one has not yet been able to find a simultaneous accounting of
both the parity conserving P-wave and parity violating (PV) S-wave decay amplitudes. Similarly, the PV asymmetries associated with the radiative
decays of hyperons are anomalously large. In the limit of degenerate u-, d-, and s-quarks, SU(3) flavor symmetry implies that these asymmetries
must vanish. Given the known mass splitting between the strange and two light flavors, one would expect the asymmetries to  have magnitudes of
order $m_s/M_B\sim 0.15$, where $M_B\sim 1$ GeV is a typical hyperon mass. The experimental asymmetries, in contrast, are four-to-five times
larger in magnitude. Even the well-known $\Delta I=1/2$ rule that summarizes the observed dominance of the $I=1/2$ channel over the $I=3/2$
channel in strangeness changing  nonleptonic decays remains enigmatic, as no apparent symmetry favors either channel. In short, consideration of
QCD symmetries and the relevant physical scales does not suffice to account for the observed properties of the $\Delta S=1$ HWI.

While the puzzles surrounding the strangeness changing HWI have been discussed extensively elsewhere, the $\Delta S=0$ HWI has generally
received less attention. Nonetheless, since we do not know whether the breakdown of QCD symmetry-based expectations in the $\Delta S=1$ sector
results from the presence of a dynamical strange quark or from other, yet-to-be-uncovered dynamics, consideration of the $\Delta S=0$ HWI -- for
which the strange quark plays a relatively minor role -- is no less important. In the following review, we focus on this component of the HWI.

According to the SM, the structure of the low-energy $\Delta S=0$ HWI is relatively simple: \be \label{eq:hwiquarks} {\cal H}_{\rm HWI}^{\Delta
S=0} = \frac{G_F}{\sqrt{2}}\left(J_\lambda^{CC\, \dag} J^{\lambda\, CC} +\frac{1}{2}J_\lambda^{NC\, \dag} J^{\lambda\, NC}\right)\ \ \ \, \ee
where $G_F$ is the Fermi constant and where $J_\lambda^{CC}$ and $J_\lambda^{NC}$ are the weak charged and neutral currents, respectively.  The
theoretical challenge is to find the appropriate effective interaction ${\cal H}_{\rm HWI}^{\Delta S=0\ \rm eff}(N,\pi,\Delta, \ldots)$ that
best describes the hadronic manifestation of ${\cal H}_{\rm HWI}^{\Delta S=0}$. Because $J_\lambda^{CC}$ transforms as a doublet under strong
isospin while $J_\lambda^{NC}$ contains $I=0$ and $I=1$ components, the current-current products in ${\cal H}_{\rm HWI}^{\Delta S=0}$ contain
terms that transform as isoscalars, isovectors, and isotensors.  Consequently, ${\cal H}_{\rm HWI}^{\Delta S=0\ \rm eff}$ must contain the most
general set of operators having the same isospin properties. In what follows, we review the theoretical efforts to determine this effective
interaction.

Experimentally, the $\Delta S=0$ HWI can be isolated solely  via hadronic and nuclear physics processes that violate parity, thereby filtering
out the much larger effects of the strangeness conserving strong and electromagnetic  interactions. Efforts to do so are not new. Soon after the
1957 discovery of parity violation in $\mu$-decay and nuclear $\beta$-decay, the search was on for evidence of a PV weak nuclear force that
would result in small, parity violating effects in nuclear observables. That year, Tanner reported the first experimental search for a PV
nucleon-nucleon (NN) interaction \cite{tanner}. Subsequently, Feynman and Gell-Mann \cite{Feynman:1958ty} predicted that the four fermion
interactions responsible for leptonic and semi-leptonic weak decays should have a four nucleon partner that is similarly first order in $G_F$. A
decade later, Lobashov et al.\ produced the first definitive evidence for the existence of a first order weak NN force in radiative neutron
capture on $^{181}$Ta that was consistent with the Feynman and Gell-Mann hypothesis \cite{lobashov1, lobashov2}.

The pursuit of this evidence in the Tanner, Lobashov and subsequent experiments was challenging, as one expected the magnitude of the PV effects
to be ${\cal O}(10^{-7})$. Along the way, it was realized that certain accidents of nuclear structure in many-body nuclei could amplify the
expected PV effects by several orders of magnitude, and a $\sim 10\%$ PV effect was, indeed, observed in $^{139}$La \cite{bowman:1989ci}.  The
amplification arises from two sources: the presence of nearly degenerate opposite parity states that are mixed by the HWI, and the interference
of an otherwise parity forbidden transition amplitude with a much larger parity allowed one. Subsequent experiments then yielded a mix of  PV
measurements in nuclei, where one expected amplification factors of order $10^2$ to $10^3$, as well studies of PV observables in the scattering
of polarized protons and neutrons from hadronic targets.

Theoretically, however, the use of nuclear systems introduces an additional level of complication in the interpretation of experiments, as one
must contend with both nuclear structure effects as well as the dynamics of  non-perturbative QCD.  For over two decades now, the conventional
framework for carrying out this interpretation has been a meson exchange model, popularized by the seminal work of Desplanques, Donoghue, and
Holstein (DDH) \cite{Desplanques:1979hn}. The model assumes that the PV nucleon-nucleon (NN) interaction is dominated by the exchange of the
pion and two lightest vector mesons ($\rho$ and $\omega$), and its strength is characterized by seven PV meson-nucleon couplings: $\hpi$,
$h_\rho^{0,1,2}$, $\hrhoop$ and $h_\omega^{0,1}$, where the superscript indicates the isospin\footnote{In the literature, the isovector, PV $\pi
NN$ coupling is often denoted $f_\pi$. Here, however, we adopt the $\hpi$ notation to avoid confusion with the pion decay constant.}. DDH
provided theoretical \lq\lq reasonable ranges"  and \lq\lq best values" for the $h_M^i$ using SU(6) symmetry, constraints from non-leptonic
hyperon decay data, and the quark model to estimate the experimentally unconstrained terms. Despite various attempts to improve upon the
original DDH work, the results of their analysis still remain as the benchmark, theoretical targets for the PV meson-nucleon couplings.

The experimental results from nuclear and hadronic PV measurements have been analyzed using the DDH framework, leading to constraints on
combinations of the $h_M^i$ that typically enter PV observables. The results are in general agreement with the DDH reasonable ranges, though the
ranges themselves are quite broad, and the constraints from different experiments are not entirely consistent with each other. A particular
quandary involves $\hpi$: the $\gamma$-decays of $^{18}$F imply that it is consistent with zero, while the analysis of the $^{133}$Cs anapole
moment differs from zero by several standard deviations \cite{Haxton:2001zq}. More to the point, the connection between the PV experiments and
SM expectations is far from transparent. Indeed, in order to  draw this connection using the meson-exchange framework and nuclear PV
observables, one has to sort through a number of model dependent effects involving nuclear structure, hadron structure, and the meson exchange
model itself. Whether one has a reasonable hope for doing so in a systematic manner is debatable at best.

\begin{figure}[h]
\centerline{\epsfig{figure=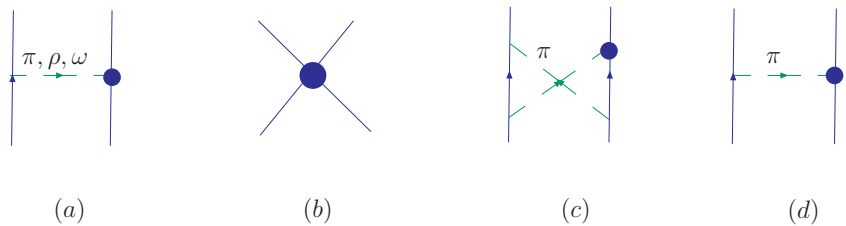,width=4.75in}} \caption{{\em Comparison of (a) meson-exchange and (b-d) effective field theory (EFT)
treatments of the parity-violating NN interaction. Panels (b), (c), and (d) give illustrative contributions to short, medium, and long-range
components, respectively .}} \label{fig:mexvseft}
\end{figure}

At the end of the day, the goal of studying the $\Delta S=0$ HWI with hadronic and nuclear PV is to help determine the degree to which the
symmetries of QCD characterize the realization of the HWI in strongly interacting systems and, as a corollary, to shed light on the long
standing puzzles in the $\Delta S=1$ sector. To that end, one would ideally formulate the problem to make the contact with the underlying SM as
transparent as possible while avoiding hadronic model and nuclear structure ambiguities. Recently, a framework for doing so has been formulated
in Reference \cite{Zhu:2004vw} using effective field theory ideas. That work builds on the extensive developments in the past decade of an EFT
for the strong NN interaction that has been applied successfully to a variety of few-body nuclear phenomena. In the case of the PV NN force, two
versions of the EFT are useful, depending on the energy scales present in the process under consideration:

\begin{itemize}

\item[(I)]
For energies well below the pion mass, the EFT contains
only four-nucleon operators and five effective parameters, or
\lq\lq low-energy constants",  that characterize the five
independent low-energy S-wave/P-wave mixing matrix elements:
$\lambda^{0,1,2}_s$, $\lambda_t$, and $\rho_t$. Relative to the
leading order parity-conserving four nucleon operators, the PV
operators are ${\cal O}(Q)$, where $Q$ is a small energy scale. In
this version of the EFT, the pion is considered to be heavy and
does not appear as an explicit, dynamical degree of freedom.

\item[(II)] At higher energies, the the pion becomes dynamical and three additional constants associated with $\pi$-exchange effects appear at
lowest order: $\hpi$,  along with a second parameter in the EFT potential, $k_\pi^{1a}$, and a new meson-exchange current operator characterized
by ${\bar C}_\pi$. Moreover, the EFT incorporates the effects of two-pion exchange for the first time in a systematic way, leading to
predictions for a medium-range component of the PV NN interaction.

 \end{itemize}
The essential differences between the PV EFT and the meson-exchange frameworks -- as well as their similarities -- are summarized in Figure
\ref{fig:mexvseft}.

Clearly, implementing the EFT approach to the $\Delta S=0$ HWI requires carrying out new experiments in few-body systems for which ab initio
structure computations can be performed. As outlined in Reference~\cite{Zhu:2004vw}, a program of such measurements exists in principle. From a
practical standpoint, carrying it out will involve meeting a number of experimental challenges. In light of these new theoretical developments
and experimental opportunities, we believe it is time to review the field of hadronic PV anew. Comprehensive reviews of the subject have
appeared over the years, including the influential Annual Reviews article by Aldelberger and Haxton completed two decades ago
\cite{Adelberger:1985ik}. In what follows, we hope to provide the \lq\lq next generation" successor to that work, updating the authors' analysis
in light of new theoretical and experimental progress. Since our focus will be on new developments, we touch only lightly on older work that has
been reviewed in Reference~\cite{Adelberger:1985ik} and elsewhere \cite{Desplanques:1998ak}. Before doing so in detail, however, we find it
useful to summarize the primary developments and shifts in emphasis that have occurred since Reference \cite{Adelberger:1985ik} appeared:

\begin{itemize}

\item The extensive development of $\chi$PT and NN EFT, together
with substantial progress in performing lattice QCD simulations,
has revolutionized our approach to treating hadronic physics.
While the use of hadronic models can provide important physical
insights, the present day \lq\lq holy grail" is to derive
first-principles QCD predictions for hadronic phenomena. At the
time of the Adelberger and Haxton review, the quark model was
still in vogue, whereas lattice QCD and hadronic EFTs had yet to
realize their potential. Today, the situation is reversed. Indeed,
in the case of $\Delta S=0$ HWI, the use of a meson-exchange model
for the NN interaction that entails a truncation of the QCD
spectrum and contains effective couplings that likely parameterize
more physics than the elementary meson-nucleon PV interaction
({\em e.g.}, $2\pi$-exchange) obscures rather than clarifies the
connection with the SM. We now know how to do better.

\item New experimental and technological developments have opened the way to performing PV experiments in few-body systems. The landscape now
differs substantially from  that of the 1980's, at which time it appeared that measuring a number of ${\cal O}(10^{-7})$ effects in few-body
processes was impractical. Indeed, two decades ago, the presence of the nuclear enhancement factors made experiments with many-body nuclei such
as $^{18}$F more attractive than those in  few-body systems. Since then, precise new measurements of $10^{-7}$ PV observables in ${\vec p}p$
scattering, ${\vec n}\alpha$ spin rotation, and polarized neutron capture on hydrogen have either been completed or are in progress, and plans
are being developed for other similarly precise few-body measurements at NIST, LANSCE, the SNS, and IASA (Athens). As we discuss below,
completion of a comprehensive program of few-body measurements is now a realistic prospect.

\item Enormous progress has been made in performing precise,  ab initio calculations in the few-body system using Green's function and
variational Monte Carlo methods. These computations start with state-of-the-art phenomenological potentials that incorporate our present
knowledge of NN phase shifts and include minimal three-body forces as needed to reproduce the triton binding energy and other three-body
effects. A marriage between the NN EFT methods and these few-body computational approaches is also being developed. As a result, a realistic
prospect exists for performing precise computations with the PV EFT for few-body observables, leaving one free from the nuclear structure
questions that enter the interpretation of many-body PV observables.

\end{itemize}

In short, the frontier today for understanding the $\Delta S=0$
HWI lies in the few-body arena, for which a combination of precise
experiments and first-principles theory provide new tools for
making the most direct possible confrontation with the interplay
of the strong and electroweak sectors of the SM. In the remainder
of this article, we elaborate on this view.

\section{Weak Meson Exchange Model Meets the End of the Road}

While the era of the meson-exchange framework for hadronic PV is drawing to a close, it has played such a central role in the field that its
development and use following the publication of Reference~\cite{Adelberger:1985ik} calls for a brief review.  The primary theoretical
developments have included updated theoretical  \lq\lq reasonable ranges" and \lq\lq best values"  for the $h_M^i$ provided by DDH and others
\cite{Desplanques:1979hn, Dubovik:1986pj, Feldman:1991tj}, the analysis of nuclear anapole moments extracted from atomic PV experiments,
computations of nuclear PV contributions to PV electron scattering asymmetries, and new global fits of the $h_M^i$ to nuclear and hadronic PV
data. Experimentally, one has seen the completion of the TRIUMF 221 MeV ${\vec p} p$ scattering experiment and a neutron spin rotation
experiment at NIST, the launching of an ${\vec n}p\to d\gamma$ experiment at LANCSE, and the first non-zero result for a nuclear anapole moment
in an atomic PV experiment with $^{133}$Cs.

\subsection{Meson Exchange Model of the Weak N-N Interaction}
\label{subsec:mecmodel}

The meson-exchange, PV NN potential, $V^{\rm PV}_{\rm DDH}$, is generated by the meson-exchange diagrams of Figure \ref{fig:mexvseft}a, wherein
one meson-nucleon vertex is  parity conserving and the other parity violating. The Lagrangians for each set of interactions have been written
down on numerous occasions in the literature, so we only give the final form of the static potential:
\begin{eqnarray}
V^{\rm PV}_{DDH}(\vec{r}) &=&i{\hpi g_{A}m_N\over
\sqrt{2}F_\pi}\left({\tau_1\times\tau_2\over 2}\right)_3
(\vec{\sigma}_1+\vec{\sigma}_2)\cdot
\left[{{\vec p}_1-{\vec p}_2\over 2m_N},w_\pi (r)\right]\nonumber\\
&&-g_\rho\left(h_\rho^0\tau_1\cdot\tau_2+h_\rho^1\left({\tau_1+\tau_2\over
2} \right)_3+h_\rho^2{(3\tau_1^ 3\tau_2^3-\tau_1\cdot\tau_2)\over
2\sqrt{6}}\right)
\nonumber\\
&&\quad \left((\vec{\sigma}_1-\vec{\sigma}_2)\cdot \left\{{{\vec
p}_1-{\vec p}_2\over 2m_N},w_\rho(r)\right\}
+i(1+\chi_\rho)\vec{\sigma}_1\times\vec{\sigma}_2\cdot
\left[{{\vec p}_1-{\vec p}_2\over 2m_N},w_\rho
(r)\right]\right)\nonumber\\
&&-g_\omega\left(h_\omega^0+h_\omega^1\left({\tau_1+\tau_2\over
2}\right)_3\right)\nonumber\\
&&\quad\left((\vec{\sigma}_1-\vec{\sigma}_2)\cdot \left\{{{\vec
p}_1-{\vec p}_2\over 2m_N},w_\omega (r)\right\}
+i(1+\chi_\omega)\vec{\sigma}_1\times\vec{\sigma}_2\cdot
\left[{{\vec p}_1-{\vec p}_2\over
2m_N},w_\omega(r)\right]\right)\nonumber\\
&&-\left(g_\omega h_\omega^1-g_\rho h_\rho^1\right)
\left({\tau_1-\tau_2\over 2}\right)_3
(\vec{\sigma}_1+\vec{\sigma}_2)\cdot \left\{{{\vec p}_1-{\vec
p}_2\over 2m_N},w_\rho(r)\right\}
\nonumber\\
\label{eq:DDH1} &&-g_\rho
h_\rho^{'1}i\left({\tau_1\times\tau_2\over 2}\right)_3
(\vec{\sigma}_1+\vec{\sigma}_2)\cdot \left[{{\vec p}_1-{\vec
p}_2\over 2m_N},w_\rho(r)\right] \ \ \ .
\end{eqnarray}
Here ${\vec p}_i=-i{\vec\nabla}_i$, with ${\vec\nabla}_i$ denoting the gradient with respect to the coordinate ${\vec x}_i$ of the $i$-th
nucleon,
 $r=|{\vec x}_1 - {\vec x}_2|$ is the separation between the two
nucleons,
\begin{equation}
w_i(r)=\frac{\exp (-m_ir)}{4\pi r} \label{wi(r)}
\end{equation}
is the standard Yukawa function, and the strong $\pi NN$ coupling $g_{\pi NN}$ has been expressed in terms of the axial-current coupling $g_A$
using the Goldberger-Treiman relation: $g_{\pi NN}= g_A m_N/F_\pi$, with $F_\pi =92.4$ MeV being the pion decay constant. The $g_V$,
$V=\rho,\omega$, are the strong vector meson-nucleon Dirac couplings, and the $\chi_V$ give the ratio of the strong Pauli and Dirac couplings.
The terms in Eq. (\ref{eq:DDH1}) display different dependences  on isospin and spin, so that various observables are sensitive to distinct
linear combinations of the $h_M^i$. A notable feature is the absence of a neutral $\pi$-exchange component. Indeed, the only manifestation of
$\pi$-exchange appears in the first term of Eq. (\ref{eq:DDH1}) that contains only products of the isospin raising and lowering operators for
the two nucleons. This feature reflects a more general theorem by Barton that forbids a neutral pseudoscalar-exchange component in the PV
potential when CP is conserved \cite{Barton:1961eg}.

The values of the $h_M^i$ appearing in $V^{\rm PV}_{\rm DDH}$ are most conveniently expressed  in units of $g_\pi$, the natural strength for the
weak $\Delta S=1$ $B\to B^\prime \pi$ couplings\footnote{Here, $B$ and $B'$ denote octet baryons.}: \be g_\pi = 3.8\times 10^{-8} \approx
\frac{G_F F_\pi^2}{2\sqrt{2}} \ \ \ .  \label{eq:gpidef} \ee The original DDH reasonable ranges and updated best values are given in Table
\ref{tab:DDH}.  Note that no prediction for $h_\rho^{1 \prime}$ appears, as DDH were unable to compute this constant in
Reference~\cite{Desplanques:1979hn}. Subsequently, Holstein \cite{Holstein:1981cg} used a $1/2^{-}$ pole model to estimate this parameter. Using
the quark model to compute the $1/2^{-} \leftrightarrow 1/2^{+}$ mixing matrix elements, he obtained $h_\rho^{1\ \prime}\simeq 1.8 \; g_\pi$.
Henceforth, we will not refer to this prediction when referring to the DDH values.

\begin{table}
\def~{\hphantom{0}}
\caption{{\em Theoretical reasonable ranges (second column) and best values (columns 3-5) for the PV meson-nucleon couplings
\cite{Holstein:2004kx}, $h_M^i$, from DDH \cite{Desplanques:1979hn}, Dubovic and Zenkin (DZ)\cite{Dubovik:1986pj}, and Feldman et al.\
\cite{Feldman:1991tj}.  All values are quoted in units of $g_\pi = 3.8 \times 10^{-8}$. \vspace*{0.4cm}}}\label{tab:DDH}

\begin{center}
\begin{tabular}{@{}lcccc @{}}%
%\toprule
\hline \hline PV Coupling & DDH  Range & DDH   Best Value & DZ  & FCDH
\\ \hline
%\colrule
$h^1_\pi$        & 0 $\rightarrow$  30      & + 12  & +3    & +7     \\
$h^0_\rho$       & 30 $\rightarrow$  -81   & -30   & -22   &   -10    \\
$h^1_\rho$       & -1 $\rightarrow$ 0       & -0.5  & +1    &   -1     \\
$h^2_\rho$       &  -20 $\rightarrow$ -29   & -25   & -18   &  -18  \\
$h^0_\omega$     & 15 $\rightarrow$ -27     & -5    & -10   &   -13     \\
$h^1_\omega$     & -5 $\rightarrow$ -2      & -3    & -6    &   -6     \\

%\botrule
\hline
\end{tabular}
\end{center}
\end{table}

%   --------
%    below is the format for the ANNREV style files. comment it out for now.
%\begin{table}
%\def~{\hphantom{0}}
%\caption{\label{tab:DDH} This table is taken from Holstein, 2004.
%All values are quoted in units of $g_\pi = 3.8 \times 10^{-8}$. }
%\begin{tabular}{@{}lcccc@{}}%
%\toprule
%Weak M-N Coupling & DDH \cite{DDH80} Range & DDH \cite{DDH80}  Best Value & DZ \cite{Dubovic} & FCDH  \cite{Feldman} \\
%\colrule
%$f^1_\pi$        & 0 $\rightarrow$  30      & + 12  & +3    & +7     \\
%$h^0_\rho$       & 30 -$\rightarrow$  -81   & -30   & -22   &   -10    \\
%$h^1_\rho$       & -1 $\rightarrow$ 0       & -0.5  & +1    &   -1     \\
%$h^2_\rho$       &  -20 $\rightarrow$ -29   & -25   & -18   &  -18  \\
%$h^0_\omega$     & 15 $\rightarrow$ -27     & -5    & -10   &   -13     \\
%$h^1_\omega$     & -5 $\rightarrow$ -2      & -3    & -6    &   -6     \\
%\botrule
%\end{tabular}
%\end{table}
%---------------------------------------------------
The various SU(6)$_w$ symmetry arguments, current algebra techniques, and quark model estimates that lead to the values in Table \ref{tab:DDH}
have been discussed in detail elsewhere \cite{Desplanques:1979hn, Feldman:1991tj}, and since our emphasis lies on a new formulation in which
these couplings do not
appear, we do not revisit those discussions here. Instead, we concentrate on new applications of this framework.\\

\noindent {\bf Anapole Effects}

Two particularly novel uses of the PV meson-exchange framework have been in the analysis of atomic PV experiments and PV electron scattering.
Shortly after PV was observed in $\mu$-decay and $\beta$-decay, Zeldovich and Vaks pointed out that weak interactions could also induce a PV
coupling of the photon and fermion \cite{Zel57}. Electromagnetic (EM) gauge invariance implies that the lowest dimension effective operator for
this coupling has the form \cite{Musolf:1990sa} \be \label{eq:ana1} {\cal L}_{\rm PV}^{ff\gamma} = \frac{F_A}{\Lambda^2}\ {\bar\psi}_f
\gamma_\mu\gamma_5\psi_f
\partial_\nu F^{\mu\nu}\ \ \ ,
\ee
where $F^{\mu\nu}$ is the EM field strength tensor, $F_A$ is
the anapole coupling, and $\Lambda$ is an appropriate mass scale.
This effective operator leads to the momentum-space interaction
\be \label{eq:ana2} M_{\rm PV}^{eff} = -\frac{F_A}{\Lambda^2}\
{\bar u}(p) \left[ q^2\gamma_\mu-\not\! q q_\mu\right]\gamma_5
u(p) \varepsilon^\mu(q)\ \ \ , \ee where $q=p'-p$ is the momentum
of a photon with polarization vector $\varepsilon^\mu(q)$. From
(\ref{eq:ana2}), it is clear that the anapole coupling involves
only virtual photons, since for a real photon $q^2=0$ and one can
always choose a gauge in which $q\cdot\varepsilon=0$. Since $\partial_\nu F^{\mu\nu}=J^\mu$,
Eq. (\ref{eq:ana1}) implies that the anapole interaction effectively
couples the fermion axial current to the source of the EM field,
$J^\mu$.

In the 1980's, Flambaum, Khriplovich, and Sushkov \cite{Flambaum:1980sb,Flambaum:1984fb} observed that the anapole moments of nuclei would scale
as the square of the nuclear radius, rather than as $1/\Lambda^2$, so that their magnitudes would be enhanced as $A^{2/3}$ in heavy nuclei.
Moreover, the nuclear anapole moment would couple the nuclear axial current to the EM currents of the atomic electrons, thereby inducing a PV,
nuclear spin-dependent (NSD) term in the atomic Hamiltonian. Experimentally, one could isolate this effect by observing NSD transitions in
atomic PV processes. As discussed below, a non-zero result for the $^{133}$Cs anapole moment has been obtained by the Boulder group, while
limits have been placed on the anapole moment of $^{205}$Tl by the Seattle group. Efforts are presently underway to measure the anapole moments
of other nuclei, such as francium (for recent reviews, see e.g. references~\cite{Haxton:2001ay, Dmitriev:2003hs, Ginges:2003qt}).

\begin{figure}[h]
\vspace*{0.7cm} \centerline{\epsfig{figure=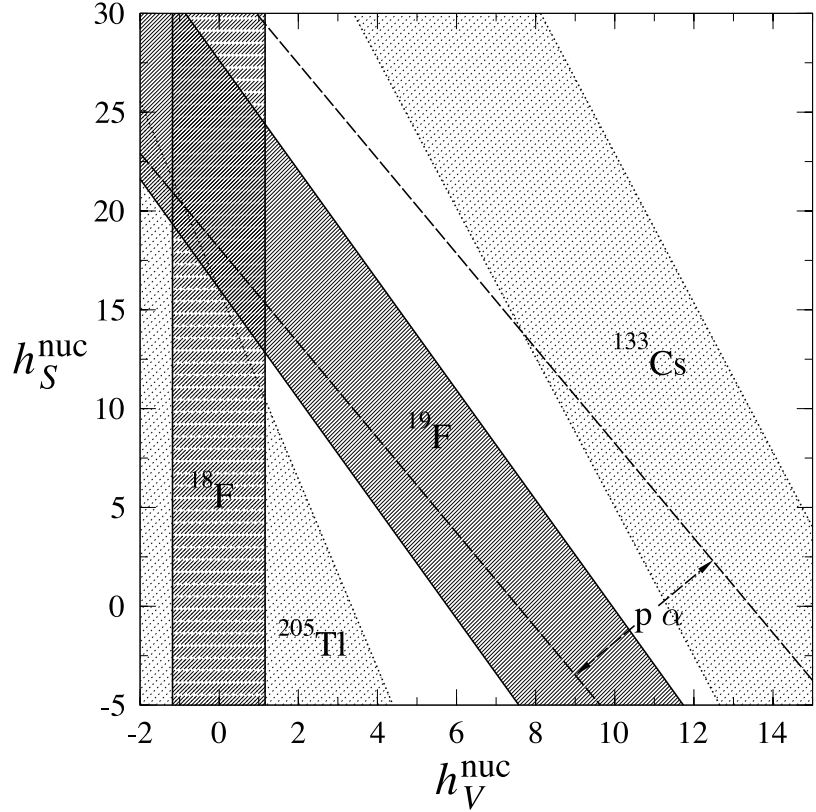,width=4.2in}}  \caption{{\em Constraints on effective DDH weak meson-nucleon couplings
deduced from PV observables in nuclei and anapole moments of heavy atoms (courtesy of W.~C.~Haxton). Here, $h_V^{\rm nuc}=h_\pi^1-0.12
h_\rho^1-0.18 h_\omega^1$ and $h_S^{\rm nuc}= - (h_\rho^0+0.7 h_\omega^0$) } }\label{nuclearlimits}
\end{figure}

The new anapole moment measurements have stimulated considerable theoretical activity. Using a one-body averaged version of $V^{\rm PV}_{\rm
DDH}$ and a simple single particle shell model, the authors of references ~\cite{Flambaum:1980sb,Flambaum:1984fb} estimated the magnitude of the
anapole moments of various nuclei, demonstrating the $A^{2/3}$ scaling under these conditions. Substantially more sophisticated shell model
calculations -- using the complete two-body potential and associated meson-exchange currents -- were carried out in references
\cite{Haxton:1989ap,Haxton:2001mi,Haxton:2001zq}. The results have been used to extract constraints on the DDH couplings, as shown in Figure
\ref{nuclearlimits} .  Of particular interest is the significant disagreement between the $^{133}$Cs anapole constraints on the relevant
isovector combination of couplings, compared with those obtained from the circular polarization $P_\gamma$ in the $\gamma$-decay of $^{18}$F.
Other nuclear model computations of the $^{133}$Cs anapole moment, having greater or lesser degrees of sophistication, lead to similar
conclusions \cite{Bouchiat:1990bu, Bouchiat:1991hw, Dmitriev:1994ct, Dmitriev:1999xq, Wilburn:1998xq, Auerbach:1999jb}.

As discussed in references ~\cite{Haxton:2001mi,Haxton:2001zq}, state-of-the art shell model computations unavoidably entail model-space
truncations, and in the case of the cesium anapole moment calculations, inclusion of the omitted contributions would likely increase, rather
than decrease, the disagreement with the $^{18}$F result. In contrast, the nuclear structure analysis used to interpret the $P_\gamma$ results
is thought to be robust, since the dominant nuclear mixing matrix element can be calibrated against an analog $\beta$-decay amplitude
\cite{bennett80,Haxton:1981sf}. Thus, the implications of the new result for the Cs anapole moment are quite puzzling.

\begin{figure}[h]
\centerline{\epsfig{figure=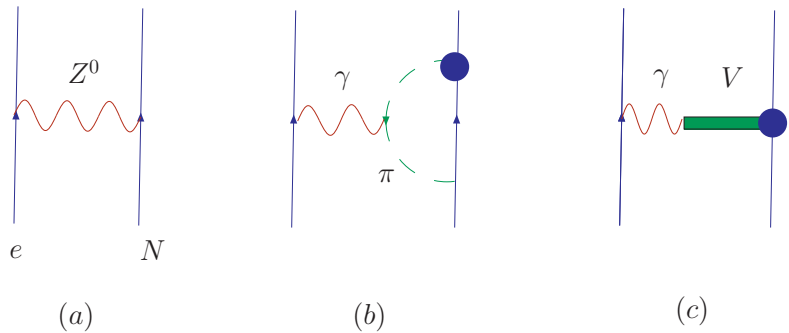,width=4.75in}} \caption{{\em Contributions from hadronic PV (proton anapole moment)  to the effective
axial vector electron-proton coupling, $G_A^e$.} }\label{fig:gae}
\end{figure}

The $\Delta S=0$ HWI can also contribute to the PV asymmetry $A_{\rm PV}$  measured in the scattering of longitudinally polarized electrons from
hadronic and nuclear targets. In the late 1990's, it was realized that the anapole moment of the proton contributes to $A_{\rm PV}$ for elastic
${\vec e}p$ scattering in a way that is indistinguishable from that of the axial vector coupling of the $Z^0$ to the proton, or $G_A^e$ (see
Figure \ref{fig:gae}) \cite{Musolf:1990ts}. Moreover, it was shown -- using the DDH framework -- that the both the magnitude of the proton
anapole moment contribution as well as the theoretical uncertainty associated with it was sufficiently large as to significantly affect the
interpretation of $A_{\rm PV}$. At that time, a program of PV electron scattering measurements was being developed to determine the strange
quark contributions to the electric and magnetic form factors of the proton, $G_E^s$ and $G_M^s$, respectively. The presence of the anapole
related uncertainties would be particularly problematic for the extraction of $G_M^s$ from the backward angle asymmetry measurements
\cite{Musolf:1992xm,Musolf:1993tb}.

Consequently, additional measurements of $A_{\rm PV}$ for quasielastic (QE) scattering from the deuteron were carried out. Since the deuterium
asymmetry is strongly sensitive to $G_A^e$ but considerably less sensitive to $G_M^s$ than is the proton asymmetry, a measurement of $A_{\rm
PV}^{\rm QE}({\vec e}D)$ -- in conjunction with $A_{\rm PV}^{\rm El}({\vec e}p)$ -- could be used to test the theoretical estimates of
Reference~\cite{Musolf:1990ts} while providing for a determination of $G_M^s$ that is independent of hadronic PV uncertainties. The initial
results of these measurements, completed by the SAMPLE Collaboration \cite{Hasty:2001ep}, yielded a new puzzle: the effective $G_A^e$ extracted
from the ${\vec e}p$ and ${\vec e}D$ asymmetries was consistent with zero. The calculations of Reference~\cite{Musolf:1990ts} had predicted a
$\sim 40\%$ reduction to the value of $G_A^e$ arising from SM electroweak radiative corrections and the anapole effect. The SAMPLE result,
however, implied a substantial enhancement of the anapole contribution or that of other radiative corrections over the predictions of
Reference~\cite{Musolf:1990ts}.

Subsequent theoretical studies attempted to determine the origin of the anomaly, scrutinizing various contributions to $A_{\rm PV}$: the
original computation of Reference~\cite{Musolf:1990ts} was revisited and updated  using heavy baryon $\chi$PT \cite{Zhu:2000gn}; possible quark
model enhancements were considered \cite{Riska:2000qw}; the $q^2$-dependence of the anapole contribution was studied
\cite{Maekawa:2000qz,Maekawa:2000bd}, and contributions from parity mixing in the deuteron and final $np$ states were computed using the
meson-exchange model \cite{Schiavilla:2002uc,Liu:2002bq}. In all cases, no large effects were found that could resolve the puzzle. Ultimately,
the SAMPLE Collaboration carried out a reanalysis of the pion-production background in the deuterium experiment that shifted the value of
$A_{\rm PV}^{\rm QE}({\vec e}D)$ \cite{Ito:2003mr} and brought the axial term into agreement with the predictions of
references~\cite{Musolf:1990ts,Zhu:2000gn}. The theoretical results from the latter work have now been used in extracting $G_M^s$ from the
backward angle proton asymmetry \cite{Spayde:2003nr}.

Within the DDH framework, these developments -- along with the
completion of a new ${\vec p}p$ scattering experiment at TRIUMF --
have motivated reanalyses of the hadronic and nuclear PV
observables in terms of the $h_M^i$. After discussing the recent
experimental developments, we provide a summary of our current
understanding of hadronic PV in this context and make the case
that a fundamental paradigm shift is required in order to make
further progress in this field.

\subsection{Experimental Progress}
\label{sec:exptprog}

Earlier reviews, e.g. \cite{Adelberger:1985ik}, \cite{Holstein:2004kx}, \cite{Bowman:1993nu},  have documented an extensive body of experimental
work aimed at characterizing the $\Delta S = 0$ HWI,  largely carried out in many-body nuclei. As noted in the Introduction, the PV effects in
the much simpler NN and few nucleon systems were almost impractically small [${\cal O}(10^{-7})$] from the standpoint of past experimental
feasibility, so the realization of fortuitous nuclear structure effects that `amplify' the underlying NN PV signal by several orders of
magnitude naturally led to an earlier focus on many-body systems.  Future progress will hinge on a handful of  precise experiments in much
simpler few nucleon systems, for which the theoretical interpretation is less fraught with model-dependent uncertainties and for which precise
measurements now appear to be realistic. Here, we review recent experimental progress in the few-body sector and comment on new developments in
probes of PV in many-body nuclei.  These developments have been cast largely in the weak meson-exchange framework, so we will use it here in
discussion their theoretical implications. When developing the EFT framework in Section \ref{sec:EFT}, we will reframe the discussion of these
theoretical implications in the EFT formulation.

Probing the NN weak interaction in few body systems presents  significant experimental challenges.  The bare NN ${\cal O}(10^{-7})$ PV
effects compete with a host of potential systematic errors at this level, which must be both minimized through careful experimental design and
simultaneously measured to ensure that they do not obscure the true PV signal.  Even acquiring sufficient data to reach a statistical error at
the $10^{-7} - 10^{-8}$ level is no mean feat, necessitating the use of current mode detection which in turn introduces its own systematic error
sensitivities that must be controlled and understood.  The most accessible NN and few nucleon observables are accessed in polarized beam
experiments, where rapid polarization reversal provides a practical means of suppressing low frequency noise and systematic effects, at the
expense of introducing a built in sensitivity to spin correlated beam properties that can mimic the PV observable in question.   In such cases,
as much effort must be expended to optimize and characterize the polarized beam properties as is required to design and commission the PV experimental
apparatus.  Typically, a successful experiment spans a decade or more from initial concept to publication of a significant result, with
continuous refinements of the experimental apparatus and technique until the desired sensitivity is reached.  In most cases, an order of
magnitude more data are required to refine and test the apparatus than to acquire the final PV data sample.

To date, there have been a number of significant measurements of PV in $pp$ scattering, but despite several decades of experimental effort, a
definitive observation of PV in the $np$ system remains to be established. Recent experimental progress in the NN system includes the completion
of a program of high precision measurements of PV in $pp$ scattering which yield independent constraints on the weak couplings of heavier
mesons, and commissioning of a PV asymmetry measurement in ${\vec n}p\to d\gamma$ that is aimed at a precise determination of $h_\pi^1$. These
experiments will be discussed briefly below;  future possibilities for precise measurements in two and few-nucleon systems are discussed in
Section 3.

\subsubsection{Longitudinal Analyzing Power in $pp$ Scattering}

The PV observable that has been studied in $pp$ scattering is the longitudinal analyzing power, $A_z = \frac{(\sigma^+ - \sigma^-)}{ (\sigma^+ +
\sigma^-)} $, where $\sigma^+$ and $\sigma^-$ are the elastic scattering cross sections for positive and negative helicity beams incident on an
unpolarized hydrogen target. The analyzing power can be naturally expressed as a sum of parity mixed partial wave contributions, with only $S-P$
mixing required to characterize $A_z$ at low energy. The first two partial waves in the expansion are sufficient to describe $A_z$ up to several
hundred MeV; the $(^1S_0-^3P_0)$ contribution dominates at low energy, while the $(^3P_2-^1D_2)$ amplitude starts to become significant above
100 MeV. These two partial wave amplitudes have complementary dependences on the weak M-N couplings: $h^{pp}_\rho = (h^0_\rho + h^1_\rho +
h^2_\rho/\sqrt{6})$ and $h^{pp}_\omega = h^0_\omega + h^1_\omega$. As noted by Simonius \cite{Simonius:1988rg}, the dominant $S-P$ wave mixing
integrates to zero near 220 MeV beam energy, due to a fortuitous cancellation of the strong S and P wave phase shifts. This observation
motivated  the recently completed TRIUMF experiment at 221 MeV \cite{Berdoz:2002sn}, which was designed to isolate $(^3P_2-^1D_2)$ contribution.

\begin{figure}[h!]
%\epsfbox{TRIUMFapparatus.eps}
\centerline{\epsfig{figure=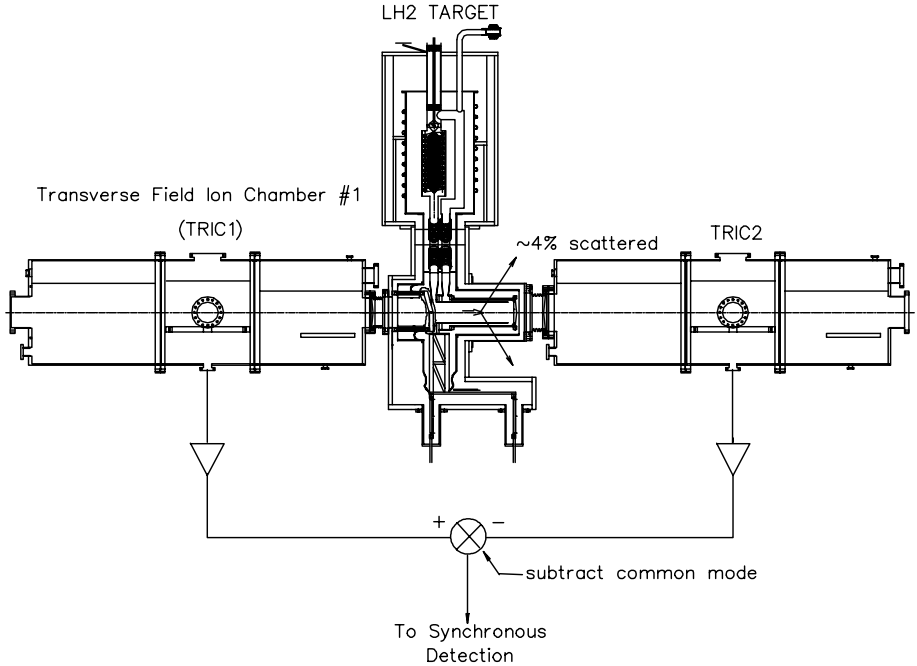,width=4.75in}} \caption{{\em Principle of the TRIUMF p-p parity violation experiment. Longitudinally
polarized protons at 221 MeV passed through a 40 cm liquid hydrogen target, which scattered $\simeq$ 4\% of the beam. The polarization-dependent
target transmission was measured by performing an analog subtraction of two dc current signals from transverse field ionization chambers.
Reprinted with permission from Berdoz et al., Phys. Rev. C Vol. 68, 034004 (2003), Fig. 1. Copyright (2003) by the American Physical Society.}}
\label{TRIUMFapparatus}
\end{figure}

While earlier high precision measurements at low energy were performed in a total scattering geometry, the TRIUMF 221 MeV measurements were
carried out in transmission mode. The small size of the total scattering asymmetry $A_z \simeq 10^{-7}$ implied a transmission asymmetry of
order 10$^{-9}$ for the TRIUMF experiment. Integrating detectors with small angular acceptance coupled to low noise electronics, excellent beam
and liquid hydrogen target stability, and a highly polarized beam with minimal helicity correlated beam properties were essential to the success
of the measurements. Rapid beam polarization reversal (40 Hz, with controlled phase slip with respect to the line frequency) led to an ac parity
violating signal at a well determined frequency, greatly suppressing the noise contributions with respect to a dc measurement (which would be
impossible in this case due to the extremely small size of the parity violating asymmetry signal). The principle of the TRIUMF transmission mode
measurement is illustrated in Figure \ref{TRIUMFapparatus}.

The TRIUMF laboratory spent many years developing a state-of-the-art optically pumped polarized H- ion source (OPPIS) \cite{OPPIS1,
Zelenski:1993nim}, which was ideal for demanding symmetry tests that require polarized beam, since the polarization ($\simeq$ 85\%) was reversed
by changing the frequency of the laser light with no changes to macroscopic electric or magnetic fields that could influence the beam
properties. Even so, significant corrections ($\simeq 40$ \%) had to be made to the raw asymmetry for helicity correlated transverse
polarization components.

Transverse polarization is a pathological source of systematic error in $pp$ scattering measurements because of the relatively enormous
parity-allowed transverse analyzing power $A_y$, which is over a million times larger than the PV longitudinal analyzing power $A_z$. While it
is not practical to demonstrate and maintain tiny transverse polarization components at or below the 10$^{-6}$ level, fortunately the
geometrical symmetry of the apparatus can be invoked to establish a `neutral axis' for the beam transport such that the false asymmetry arising
from transverse polarization components is identically zero if the beam is locked on this axis by a position feedback system. This technique was
used to ensure that corrections for average transverse polarization components $\langle P_y \rangle$ and $\langle P_x \rangle$ were
insignificant at the 10$^{-9}$ level in the TRIUMF measurements -- the associated corrections $\Delta A_z$ e.g. for transverse vertical
polarization $ \langle P_y \rangle$ scale as $ \langle x \rangle $$ \langle P_y \rangle$, where $ \langle x \rangle $ is the net displacement
from the symmetry axis, whose location can be determined by separate calibration experiments. Note that the form of the driving term $ \langle x
\rangle$$ \langle P_y \rangle$ is that of the first moment of transverse polarization at the detector location, which is referred to as an
extrinsic first moment; a similar correction must also be made for intrinsic first moments of the form $ \langle x\;P_y \rangle$ which arise
from nonuniform distributions of transverse polarization within the beam envelope. These extrinsic polarization moments are extremely difficult
to measure and control.

\begin{figure}[h]
\centerline{\epsfig{figure=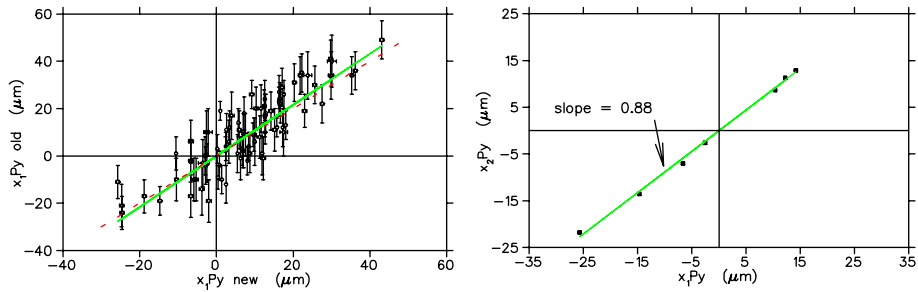,width=4.75in}} \caption{{\em Demonstration of a new current mode scanning polarimeter at TRIUMF.  Left
panel: intrinsic polarization moments $ \langle xP_y \rangle$ measured simultaneously with counting mode (vertical axis) and current mode
(horizontal axis) devices.  The errors on the current mode measurements are too small to display on this scale. The dotted line has a slope of
unity, and the solid line is a fit to the data.  Right panel: instrinsic polarization moments for a given beamline tune, measured with a pair of
current mode PPMs. The ratio of moments at PPM1 and PPM2 remains relatively constant, thus verifying an essential assumption of the systematic
error reduction scheme. Reprinted  with permission from Ramsay et al., Proc. 9th International Workshop on Polarized Sources and Targets, p.
289-293 (2002), Fig. 3.  Copyright (2002) by World Scientific. }} \label{newmoments}
\end{figure}

In the TRIUMF experiment, a pair of scanning polarimeters \cite{Berdoz:2000iz}  was employed to provide continuous measurements of the
distribution of transverse polarization components within the beam, interleaved with parity data taking from the transverse ion chambers on an
8-state, 200 ms data cycle. The limiting factor in the TRIUMF experiment was the statistical precision of the PPMs, which were coincidence
counting mode devices.  An essential assumption made in determining the correction for intrinsic polarization moments was that the moments
evolved linearly with position along the beamline and thus had a stable ratio between upstream and downstream PPMs.   This ratio could in
principle be tuned to achieve a null sensitivity to intrinsic polarization moments, an approach modelled on the successful polarization neutral
axis idea described above.  Unfortunately, the ability of the PPMs to measure this first moment ratio sufficiently precisely in a reasonable
amount of time was severely limited, and so significant corrections still had to be made to the data.

R \& D efforts towards a follow up experiment at higher energy (which was never realized) included development of a current mode scanning
polarimeter which had $\sim 20 \times$ greater statistical precision \cite{pst2001};  the current mode polarimeter was able to clearly
demonstrate the constant linear evolution of intrinsic moments along the beamline, as illustrated in figure \ref{newmoments}, thus independently
validating the corrections procedure that was used to obtain a final result for $A_z$ at 221 MeV from TRIUMF experiment 497.

\begin{figure}[h]
%\epsfbox{ppgraph.eps}
\centerline{\epsfig{figure=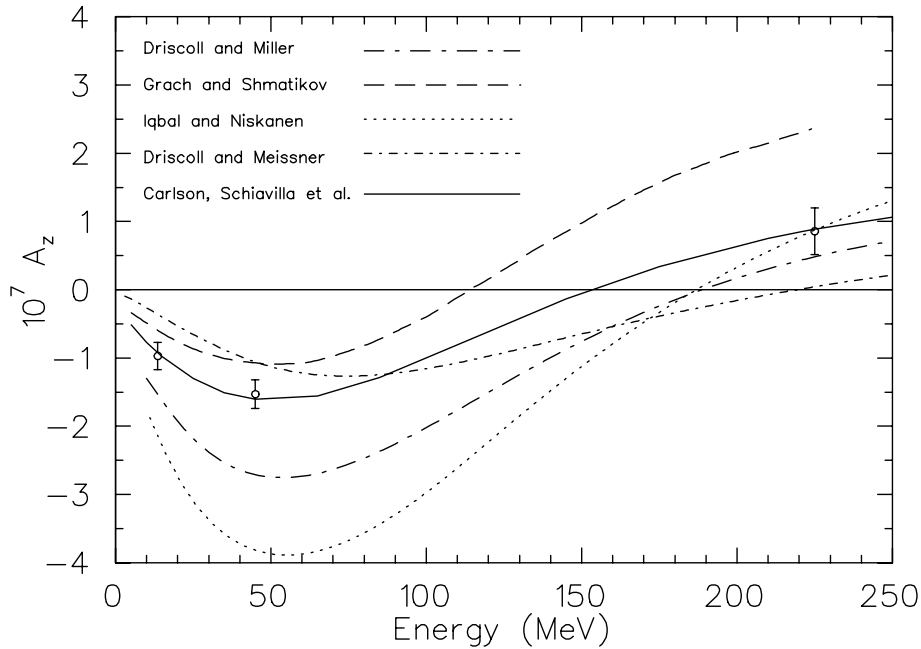,width=4.75in}} \caption{{\em The most precise measurements of parity violation in $pp$ scattering at low
and intermediate energy, and recent theoretical predictions. Experiments were performed at Bonn (13.6 MeV) \cite{Eversheim:1991tg}, PSI (45 MeV)
\cite {Kistryn:1987tq} and TRIUMF (221 MeV) \cite{Berdoz:2002sn}. The solid curve shows the calculation by Carlson et al. \cite{Carlson:2001ma}
including a fit of the weak meson-nucleon coupling constants to the data. Reprinted  with permission from Berdoz et al., Phys. Rev. C Vol. 68,
034004 (2003), Fig. 13. Copyright (2003) by the American Physical Society. }} \label{ppcalculations}
\end{figure}

Parity violation in $pp$ scattering has attracted considerable theoretical interest since the review of Reference \cite{Adelberger:1985ik}.
Recent calculations are shown in Figure \ref{ppcalculations} together with the most precise experimental data at low and intermediate energies.
Driscoll and Miller \cite{Driscoll:1989fd, Driscoll:1988hg} used the Bonn potential to treat the strong NN interaction, with weak meson-nucleon
couplings taken from Reference \cite{Desplanques:1979hn}.  Iqbal and Niskanen's calculation adds a $\Delta$ isobar contribution
\cite{Iqbal:1992xm} to the Driscoll and Miller model.  The calculation of Driscoll and Meissner \cite{Driscoll:1990ez} is based on a chiral
soliton model, while the quark model calculation of Grach and Shmatikov \cite{Grach:1993qz} takes explicit account of quark degrees of freedom.

\begin{figure}[h!]
\centerline{\epsfig{figure=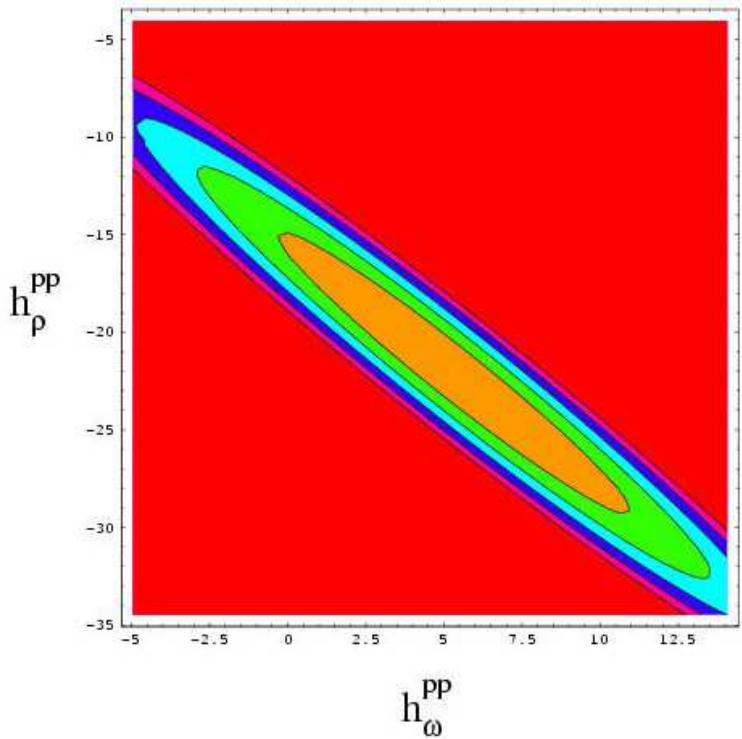,width=4.2in}} \caption{{\em Present constraints on the weak meson-nucleon couplings $h^{pp}_\rho$ and
$h^{pp}_\omega$, fitted to  the low energy and 221 MeV $pp$ asymmetry data \cite{Carlson:2001ma}. The plot shows curves of constant total
$\chi^2$ = 1,2,3,4 and 5.  Axis scales are 10$^{-7}$. Reprinted  with permission from Carlson et al., Phys. Rev. C Vol. 65, 035502 (2002), Fig.
8. Copyright (2003) by the American Physical Society.}} \label{couplings}
\end{figure}

Figure \ref{couplings} by Carlson et al.\ \cite{Carlson:2001ma} shows the limits on the weak meson-nucleon couplings $h^{pp}_\rho$ and
$h^{pp}_\omega$ imposed by the low energy $pp$ asymmetry measurements and the 221 MeV TRIUMF result. The error bands are based on a calculation
assuming the Argonne $v_{18}$ (AV-18) potential \cite{Wiringa:1994wb}, the Bonn 2000 (CD-Bonn) \cite{Machleidt:2000ge} strong interaction
coupling constants, and including all partial waves up to J=8. Although the TRIUMF measurement is not sensitive to $A_z$ from $SP$ mixing, and
the contribution from $PD$ mixing contains no $h_{\omega}^{pp}$ contribution, there is some $h_{\omega}^{pp}$ dependence arising from higher
partial waves.  The best fit to the $pp$ data yields $h^{pp}_\rho = -22.3 \times 10^{-7}$ and $h^{pp}_\omega = 5.17 \times 10^{-7}$, compared to
the DDH ``best guess'' values of $h^{pp}_\rho = -15.5 \times 10^{-7}$ and $h^{pp}_\omega = -3.0 \times 10^{-7}$.

\subsubsection{Progress in the $np$ System}
\label{sec:np}

Parity violation in the $np$ system can in principle be detected in a variety of processes that can reveal complementary aspects of the weak NN
interaction.   In $np$ capture, there are two complementary PV observables: $P_\gamma^d$, the 2.2 MeV $\gamma$ ray circular polarization for an
unpolarized neutron beam, and   $A_\gamma^d$, the asymmetry in the emission of $\gamma$ rays with respect to the neutron spin direction if the
beam is polarized. Closely related to the first of these, $P_\gamma^d$, is the helicity asymmetry $A^\gamma_L$ in the photodisintegration of
deuterium with circularly polarized photons;  the two are asymptotically equal to each other at threshold, while $A^\gamma_L$ is predicted to
drop rapidly with increasing photon energy, falling an order of magnitude as the photon energy increases to 1 MeV above threshold
\cite{Fujiwara:2004zg}. Finally, the transmission of polarized neutrons through hydrogen should reveal a tiny PV spin rotation about the neutron
propagation direction $\hat{z}$: $d\phi^{np}/dz$.

Unfortunately, all of these $np$ system measurements are extremely challenging; the first three have been attempted \cite{knyazkov84,
Earle:1988fc, Cavaignac:1977uk} but have yielded null results with limits at least one order of magnitude too large to provide a meaningful
constraint on the weak meson-exchange predictions \cite{Oka:1983sp, Schiavilla:2004wn, Khriplovich:2000mb, Liu:2004zm, Fujiwara:2004zg} all of
which are at the level $ 5 \times 10^{-8}$ or smaller. The two PV observables involving the $np$ capture reaction, $P_\gamma$ and $A_\gamma^d$
have complementary dependences on the weak MN couplings -- notably, $A_\gamma^d$ can yield a unique constraint on $h^1_\pi$, while $P_\gamma^d$
depends on a linear combination of $\pi$, $\rho$ and $\omega$ weak couplings \cite{Adelberger:1985ik}. Measurement of gamma ray circular
polarization requires a Compton polarimeter with typical sensitivity at the few percent level, thus rendering the $P_\gamma^d$ measurement in
principle even more challenging than a measurement of $A_\gamma^d$.  A measurement of the helicity asymmetry $A^\gamma_L$ in the
photodisintegration of deuterium is at the early conceptual design stage  for the future experimental program at IASA, Athens
~\cite{Stiliaris:2005ab}, with the aim of reaching a sensitivity at the 10$^{-8}$ level. The $np$ spin rotation measurement, with an anticipated
$d\phi^{np}/dz$ = $ 5 \times 10^{-7}$ rad/m \cite{Carlson:2005fd} has not yet been attempted, but is envisioned as a future component of the SNS
fundamental neutron physics program, as discussed in Section \ref{sec:horizons} below.

The intrinsic interest of a clean measurement of $h^1_\pi$ in the NN system, together with considerations of experimental feasibility, have led
to the launching of a major effort at LANSCE \cite{Gericke:2005ef} -- the NPDGamma experiment -- to make a definitive measurement of
$A_\gamma^d$, with an ultimate goal of reaching $\pm$ 10\% of the DDH prediction.  The measurement of $A_\gamma^d$ requires a polarized neutron
beam with precisely known spin direction and a measurement of the 2.22 MeV gamma ray angular distribution: $\frac {d\omega}{d \theta} \sim  (1 +
A_\gamma \; cos\theta )$, where $\theta$ is the angle between the gamma ray momentum and the neutron spin.  Even at milli-eV neutron energies
for neutrons moderated in liquid hydrogen, the scattering cross section exceeds the capture cross section by a significant factor. A
parahydrogen target is essential to avoid neutron spin flip on scattering in the target, and the useful beam flux is below 15 meV to prevent
depolarization in the target.

A previous measurement of $A_\gamma^d$, performed at the ILL reactor in the 1970's \cite{Cavaignac:1977uk} reported a value of $(0.6 \pm 2.1)
\times 10^{-7}$;  the NPDGamma experiment is designed to reach an ultimate sensitivity of $\pm 5 \times 10^{-9}$, with the uncertainty dominated
by statistical rather than systematic errors. Major improvements in the experimental instrumentation and techniques that should make this
possible include:

\begin{itemize}
\item  use of a high intensity, low energy, pulsed beam, which allows for neutron energy determination via time-of-flight measurement and allows
for a separation in time of prompt $\gamma$ ray  background from the neutrons of interest;

\item  polarization of the beam via selective transmission through optically pumped polarized cells of $^3$He;  the well known spin-dependent
cross section leads to an energy dependent polarization that can be very high for low energy neutrons, and the beam polarization can be
continuously and directly monitored by online measurements of the  $^3$He cell transmission;

\item  use of a resonant RF spin flipper capable of flipping spins at all neutron energies with high efficiency and eliminating Stern-Gerlach
steering of the neutron beam associated with spin flip;

\item  implementation of a large solid angle, high efficiency CsI (Tl) gamma detector array \cite{Gericke:2004xn}
 instrumented with sensitive current mode
electronics whose intrinsic noise is negligible compared to neutron counting statistics.

\end{itemize}

\begin{figure}[hbt]
\centerline{\epsfig{figure=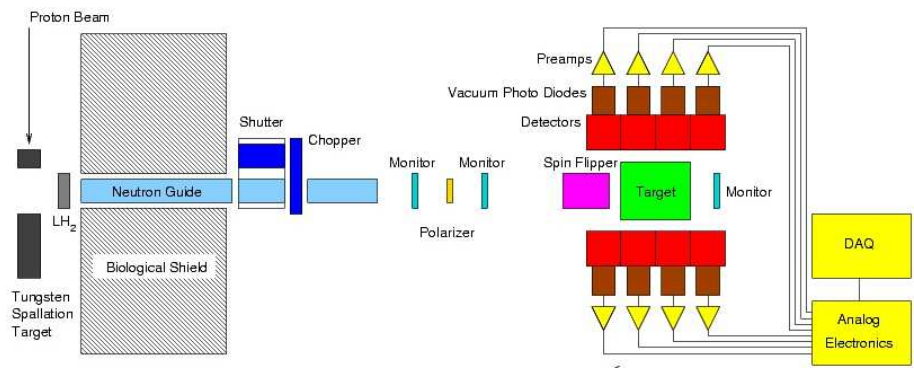,width=4.75in}} \caption{\em { Layout of the NPDGamma apparatus currently taking data on Flight Path 12 at
LANSCE. A highly uniform, vertical magnetic guide field to preserve the neutron spin direction is provided by a set of field coils (not shown).
} } \label{NPDGAMMA}
\end{figure}

\begin{figure}[hbt]
\centerline{\epsfig{figure=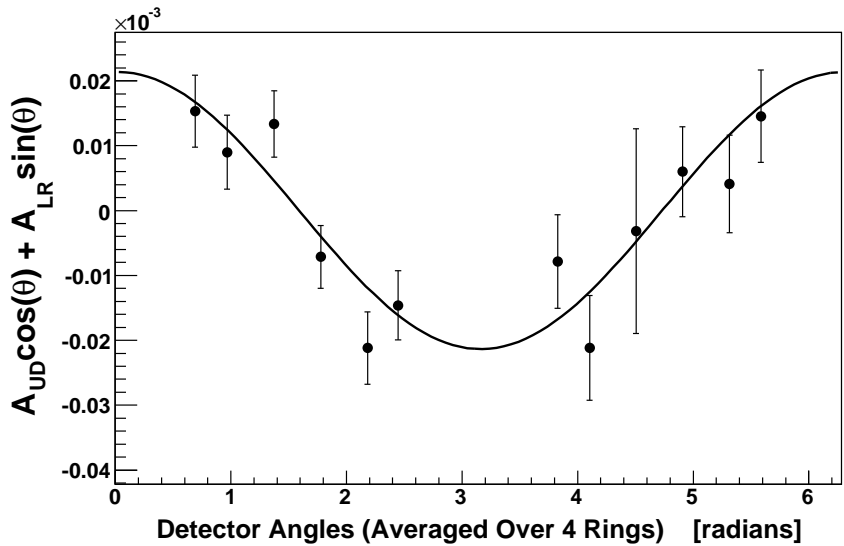,width=4.75in}} \caption{\em { CCl$_4$ gamma ray asymmetries, calculated from opposing detector pairs,
plotted versus angle of the first detector in the pair with respect to the vertical, for the NPDGamma 2004 data. The total asymmetry $A =
A_\gamma cos\theta + A_{LR} sin\theta$ is deduced from the fit. Figure courtesy of M.~T.~Gericke. }} \label{chlorine}
\end{figure}

A schematic of the NPDGamma apparatus mounted in the new experimental cave on flight path 12 at LANSCE is shown in Figure \ref{NPDGAMMA}.  The
neutron spins, polarized by the $^3$He transmission cell, are efficiently reversed by the RF spin flipper on a pulse-by-pulse basis, thus
alternating the sign of the parity violating asymmetry measured in the gamma detector array at 20 Hz. At the time of writing, NPDGamma has
achieved a number of major milestones and is ready to take production data at LANSCE, pending installation of the liquid hydrogen target. The
apparatus has been fully commissioned; electronic asymmetries are consistent with zero at the few $\times$ 10$^{-9}$ level, and PV asymmmetries
arising from neutron capture on a variety of solid targets which are representative of materials used to construct the beamline and parity
instrumentation have been measured. All are consistent with zero at the 10$^{-6}$ level or smaller, with sufficient accuracy to conclude that
background asymmetries will not play a significant role in the hydrogen target data.  A known PV asymmetry in Cl at the 10$^{-5}$ level has been
remeasured and will be used as a diagnostic tool for NPDGamma, with the periodic insertion of a CCl$_4$ target to verify the consistent
performance of the experimental setup. Contamination of the hydrogen target PV (up-down) asymmetry by the comparably small, parity-allowed
left-right asymmetry will be kept below an acceptable level by determining the effective detector angles in situ -- this in turn will be
accomplished by scanning the detector array horizontally and vertically with respect to the target while acquiring $np$ capture data.   The
measured chlorine asymmetry is shown in figure \ref{chlorine};  the net PV (up-down) and parity-allowed left-right asymmetries deduced from
these data \cite{Gericke:2005ef} are: $A_\gamma = (-19 \pm 2) \times 10^{-6}$ and $A_{LR} = (-1 \pm 2) \times 10^{-6}$.

Unfortunately, a number of factors have conspired that severely limit the statistical accuracy that can be achieved by running the NPDGamma
experiment at LANSCE.  These include a factor of 4 reduction in the available neutron flux per beam pulse as compared with expectations prior to
the recent upgrade of the  LANSCE facility, and a further reduction in high quality beam time due to magnetic interference from a neighboring
experiment. In view of these limitations, the NPDGamma collaboration plans to carry out a first measurement with hydrogen in 2005-2006, which
would provide a statistics limited result for $A_\gamma^d$ accurate to $\pm 1 \times 10^{-7}$  or better. At the time of writing, plans are
being made to move the experiment to the Fundamental Neutron Physics Beam Line at the Spallation Neutron Source, which would enable the
collaboration to make a measurement of $A_\gamma^d$ to $\pm 1 \times 10^{-8}$ or better, as discussed in Section \ref{sec:horizons}.

\subsubsection{Neutron Spin Rotation Experiments}
\label{sec:nspinrot}

Low energy neutrons can exhibit a parity violating spin rotation induced by the hadronic weak interaction as they pass through a medium.  The
observable, $\phi_{PV}$, is the angle of transverse spin rotation about the neutron's direction of motion. In the limit of zero neutron energy,
the PV neutron spin rotation is energy independent \cite{Stodolsky:1981vn}. After traversing a distance $z$ through the medium, the neutron spin
will precess by an amount $\phi_{PV} = 2 \pi \rho z f_{PV}$, where $f_{PV}$ is the parity violating coherent forward scattering amplitude for a
low energy neutron and $\rho$ is the number density of the medium \cite{Bass:2005cd}. The basic requirements of a neutron spin rotation
experiment are a source of low energy polarized neutrons, a target containing the material of interest, and a spin analyzer downstream of the
target, from which the value of $\phi_{PV}$ can be deduced.

Two cases are particularly interesting from the standpoint of testing models of the hadronic weak interaction: parity violating neutron spin
rotations in $^1$H  and $^4$He. These two cases have complementary dependences on the weak meson-nucleon couplings.   The $np$ case is dominated
by the weak pion exchange contribution, and thus would yield similar information to the NPDGamma experiment described above.  The $^4$He PV spin
rotation has a significant contribution from $h^1_\pi$ but is also sensitive to the $h^0_\rho$ coupling.  Experimentally, the $^4$He case is
more tractable, due to a much longer mean free path for low energy neutrons than hydrogen.  Like NPDGamma, the $np$ spin rotation experiment
would require a parahydrogen target to avoid neutron depolarization due to spin-exchange collisions with the hydrogen molecules, and only
neutron energies below 15 meV would be useful for the experiment.

A $^4$He spin rotation measurement was carried out at NIST in the 1990's \cite{markoff97}, with the result: $\phi_{PV} = (8 \pm 14_{stat} \pm
2_{sys} ) \times 10^{-7}$ rad/m.  Unfortunately this heroic effort did not reach sufficient sensivitity to test hadronic weak interaction
models.  A possible experiment in hydrogen has been considered but not yet proposed, and will hopefully be carried out at the SNS within the
next decade (see Section 3.3).

What confounds the beautiful simplicity of the measurement principle is the extremely small size of the PV rotations expected based on the DDH
model: for $^4$He, the predicted effect is $\phi_{PV} = -1 \times 10^{-7}$ rad/m \cite{Dmitriev83}, while for the proton, the effect should be
about 5 times larger \cite{Carlson:2005fd}. These PV spin rotations are 8 orders of magnitude smaller than those induced by the earth's magnetic
field in a typical measurement apparatus! Therefore, a much more elaborate scheme is required in order to achieve the necessary sensitivity.
Great care must be taken to reduce ambient magnetic fields by many orders of magnitude;  residual fields must be monitored carefully, and the
experiment should be designed  so that the effects of residual magnetic fields are cancelled to the maximum possible extent in the extraction of
the PV spin rotation angle.

\begin{figure}[h!]
\centerline{\epsfig{figure=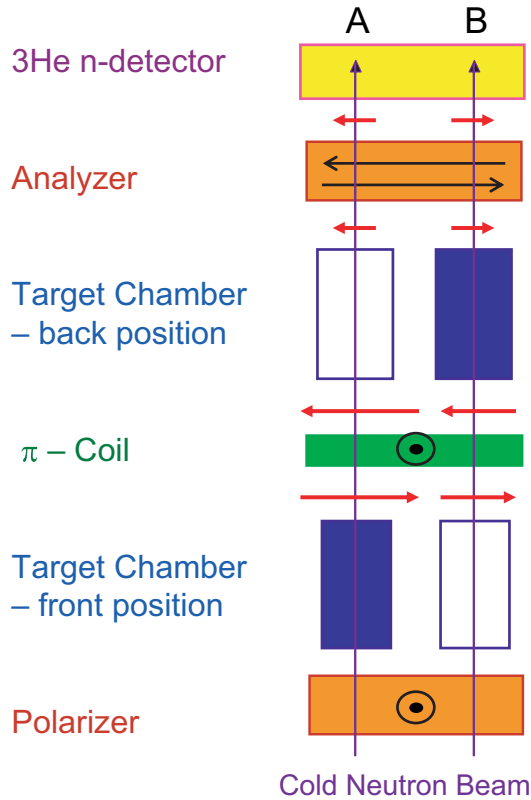,width=3.in}} \caption{\em {Schematic of the double beam / double target system for $^4$He neutron spin
rotation measurements.  The incident neutron spins are polarized out of the page. The PV effect is a spin precession around the direction of
motion, which leads to components in the horizontal plane as shown. Either the front or the back target is full (shaded), and the other is
empty, in channels A and B. Much larger spin rotations due to local magnetic fields are independent of the target state and are cancelled by
subtraction. The $\pi$ coil situated between the front and rear targets precesses the neutron spin by 180$^\circ$ about the initial spin
direction, thus reversing the sign of the PV spin rotation relative to that due to magnetic fields, indicated in the figure; the counting rate
(B - A) = 2 $\phi_{PV}$. (Figure courtesy of Anna Micherdzinska)
 }}
\label{spinrot}
\end{figure}

The basic experimental technique employed at NIST and foreseen for subsequent measurements involves a double beam / double cell apparatus, as
sketched in Figure ~\ref{spinrot}. The double cell design incorporates a 180$^\circ$ spin precession about the initial spin direction  between
the two target cells, one of which is full and one empty, for a given measurement.  A clear advantage of the double cell design is that the much
larger spin rotations due to residual magnetic fields exactly cancel in the two target states, to the extent that the magnetic fields and the
neutron trajectories are exactly the same in both cases.  An identical double target system sits beside the first one in the same cryostat, half
the beam passes through each, and the two systems are run so that at any given time, the sign of the PV spin rotation in the two subsystems is
exactly opposite.  Downstream of the double target system is a polarization analyzer and segmented $^3$He ionization chamber  which measures the
counting rate in `left' and `right' elements of the apparatus simultaneously.

Instead of rapidly flipping the beam polarization as in the $pp$  and NPDGamma experiments described earlier, the sign of the PV effect is
reversed by alternating the location of the full target cell with respect to the $\pi$ coil.  The emptying and filling of the target cells took
several minutes in the first spin rotation experiment, and each measuring interval was 10 min, corresponding to a reversal rate of 1 mHz and
roughly 25\% dead time. However, with a pair of double cell setups located side by side in the same cryostat, and a beam splitter upstream of
the apparatus, reactor beam intensity fluctuations can be effectively cancelled, which makes up for the slow PV reversal frequency.

Features of the upgraded apparatus include:  improved magnetic shielding, resulting in a fourfold reduction of ambient fields to the 20 $\mu$G
level;  improved cryogenic systems with a faster target cycling time; operation with  superfluid $^4$He,  which has a much smaller total and
small-angle neutron scattering cross section than normal $^4$He, and improvements to the NIST cold source leading to a 50\% increase in the beam
fluence.  These and other improvements should result in a statistical error of $\pm 3 \times 10^{-7}$ rad/m in a 3 month run at NIST during
2005-06, with systematic errors (based on simulations of the experiment) at the 10$^{-8}$ level.  Like NPDGamma, the collaboration proposes to
move the $^4$He spin rotation experiment to the SNS, where the high intensity pulsed cold neutron beam would permit a high statistics
measurement in about a year's running time with greater diagnostic capability for systematic errors \cite{spinrotproposal}.

\subsubsection{Nuclear Anapole Moments}

Nuclear anapole moments, which may be accessed via measurements of parity violation in atoms, have recently opened a new window on the hadronic
weak interaction \cite{Wood:1997zq}.  This subject has been the focus of significant theoretical and experimental effort, including a recent
review article in this journal \cite{Haxton:2001ay}. A detailed discussion of experimental atomic physics approaches is beyond the scope of the
present work;  however, for completeness, we include a brief overview  of  experimental efforts here.

A prime motivation for atomic PV measurements is to provide stringent tests of the electroweak Standard Model. The dominant contribution to the
neutral weak electron - nucleus interaction arises from an axial coupling to the electron and a vector coupling to the nucleus and  is
proportional to the nuclear weak charge, $Q_W$.  The value of $Q_W$ can be computed with high accuracy in the SM, and a precise experimental
determination of the nuclear weak charge can probe for deviations  that one might expect to arise from physics beyond the SM. This dominant PV
contribution is nuclear spin independent, and the extraction of $Q_W$ is minimally affected by hadronic weak interaction corrections. However,
the weak electron - nucleus interaction also has a contribution from the vector coupling of the electron to the axial current of the nucleus;
this leads to a nuclear spin dependent (NSD) parity violating interaction that can be detected as a small hyperfine dependence of atomic PV. The
parity violating interaction in this case has  contributions from $Z^0$ exchange, a hyperfine interaction correction, and finally the nuclear
anapole moment \cite{Haxton:2001mi, Haxton:2001zq}.

The weak interaction typically induces very tiny opposite-parity wavefunction admixtures in atomic states, on the order of 10$^{-11}$.  These in
turn give rise to extremely small and otherwise parity-forbidden transition amplitudes. The best explored case is atomic PV in Cs, measured in
the highly forbidden 6s-7s transition in an atomic beam, where the tiny PV amplitude was arranged to interfere with a much larger Stark-induced
transition amplitude in the presence of a static E field.\footnote{The advantage of this seemingly complicated experimental scheme is that the
interference technique yields an observable which is linear in the tiny PV amplitude, rather than trying to measure the forbidden transition
rate, which is quadratic.} The authors \cite{Wood:1997zq} report the use of 31 different servo systems to precisely control electrical, optical
and mechanical systems during the measurements;  they designed an experiment with 5 independent means of reversing the parity signal, and spent
about 20 times more data taking effort in the investigation and elimination of systematic errors compared to actual PV data taking. From the
standpoint of atomic structure calculations needed to interpret the measurements, the heavy alkali atom Cs is ideal for precision PV studies,
due to its relatively simple electronic configuration of one valence (6s) electron outside a tightly bound Xe noble gas core.

Since the nuclear anapole contribution is expected to scale as A$^{2/3}$, and the atomic PV signature used to detect it scales as
$Z^{\frac{8}{3}}$, heavy nuclei are the preferred systems for experimental studies. To date, a definitive experimental result exists only for
the anapole moment of $^{133}$Cs ~\cite{Wood:1997zq}, while an upper limit has been obtained for the anapole moment of Tl (30\% $^{203}$Tl, 70
\% $^{205}$Tl isotopic ratio) \cite{Vetter:1995vf} obtained via PV optical rotation measurements. In Cs, the nuclear spin independent atomic PV
effect has been measured to 0.4\%, while the anapole moment contribution has been determined to 14\%. In Tl, the spin independent effect has
been measured to 3\%, and the spin dependent anapole contribution is consistent with zero. These are the result of truly heroic experimental
investigations spanning over a decade of instrumentation development and testing. Interpretation of the spin independent PV as a Standard Model
test requires excellent understanding of the atomic structure, while the anapole moment must be first unravelled from the competing effects of
$Z^0$ exchange and a hyperfine correction, and then interpreted in terms of a HWI model, e.g. the meson-exchange model of DDH.

Atomic PV has been measured in other systems, but to lower precision than in Cs, and with greater uncertainty in the atomic structure details
needed to interpret the measurements.  Currently, a new generation of experiments is under development, aimed at determining both the nuclear
spin independent and anapole moment contributions to high precision. This includes ongoing efforts to develop alternative experimental
approaches to atomic PV measurements in Cs;  demonstration of a new technique based on stimulated emission has recently been reported
\cite{Guera05}, anticipating that an ultimate precision of 0.1\% could be reached. As remarked in \cite{dzuba86}, there is a significant
advantage to measuring PV in a range of isotopes of the same atom, since a number of atomic structure uncertainties cancel when PV ratios are
considered. Atomic ytterbium (Z=70), having 7 stable isotopes, has been proposed \cite{demille95} , and preparatory spectroscopic studies are
underway ~\cite{Stalnaker02}. Besides the atomic beam techniques described earlier, new experimental approaches based on trapped atoms or ions
are also being explored \cite{gomez04, Koerber03}.

Perhaps the `holy grail' of this field is the study of atomic PV in radioactive francium isotopes (Z=87), the heaviest alkali atomic system,
recently reviewed in Reference\ \cite{Gomez06}.  Both the spin independent and anapole PV contributions are expected to be roughly an order of
magnitude larger for Fr (depending on the isotope) than for Cs. In addition, Fr has a large number of isotopes spanning almost 30 neutrons with
lifetimes greater than 1 s that cover a wide range of nuclear structure conditions, which in principle permits an unprecedented systematic study
of atomic PV in a simple atomic system. With the longest lived isotope $^{223}$Fr having a half life of only 23 minutes, a radioactive beam
facility is required for a future program of Fr PV studies. Pioneering atomic spectroscopy studies of a number of trapped Francium isotopes have
been carried out at Stony Brook \cite{grossman99}, and a collaboration has recently formed with the goal of establishing a long term francium
program at the TRIUMF ISAC radioactive beam facility \cite{Gwinner05}, where an actinide production target is planned and necessary to produce
the required quantities of Francium (and other heavy systems that are of interest for fundamental symmetry tests such as proposed EDM
measurements in radon).

\subsubsection{Parity Violation in Compound Nuclei}

In contrast to the extremely small ($\simeq 10^{-7}$) PV asymmetries in the two nucleon system, many-body nuclei have over the years provided
many examples of parity violation at a much more significant scale, some of them surprisingly large (see e.g. Reference~\cite{Adelberger:1985ik}
and references therein). For the handful of cases for which reliable nuclear wavefunctions can be used to interpret the data, there is
qualitative agreement with predictions of the meson-exchange model as illustrated in Figure \ref{nuclearlimits}, albeit with a significant
discrepancy in the scale of the pion coupling as discussed earlier.  Since the publication of Reference~\cite{Adelberger:1985ik}, an extensive
and in many ways complementary program of PV measurements in compound nuclei has been carried out, dominated by the work of the TRIPLE
collaboration at LANSCE. A review of this field was presented in this journal in 1993 \cite{Bowman:1993nu} focusing on resonances in $^{238}$U
and $^{232}$Th; an updated comprehensive review of that and more recent data and analysis is provided in Reference~\cite{Mitchell:2001pr}.

The TRIPLE collaboration measured longitudinal cross section asymmetries for neutron energies in the eV - keV range on a range of nuclear
targets. Mixing of S- and P- wave resonances of the same J leads to large PV effects in some cases, with enhancement factors as large as 10$^6$
relative to the nucleon-nucleon PV asymmetries, due to the high density of states in the compound nucleus.  The measurement of a number of PV
asymmetries in the same nucleus is a key feature of this program, which is essential to the interpretation of the results.   Using a statistical
approach, the collaboration was able to extract results for either the RMS weak mixing matrix element $M_J$ or the weak spreading width
$\Gamma_w = 2 \pi M_J^2 / D_J$ where $D_J$ is the level spacing, based on a total of 75 PV resonances observed in 18 nuclei. Important
spectroscopic information (J values) for the resonant states was known in some cases, and modelled in other cases, in order to interpret the
results. The weak spreading widths were found to be on the order of 10$^{-7}$ eV and roughly constant with mass number, while typical RMS matrix
elements were on the scale of 1 meV.

Theoretical treatments have been developed that led to model predictions of the RMS matrix element based on an effective weak interaction with
dominant isovector pion and isoscalar $\rho$ exchange contributions taken from  DDH predictions.  A recent study by Tomsovic et
al.~\cite{Tomsovic:1999yb} outlines a statistical spectroscopy approach for interpreting the experimental RMS matrix elements  based on PV
asymmetry data from $^{238}$U and $^{104,105,106,108}$Pd targets to set limits on the weak meson-nucleon coupling constants.  To date, a variety
of theoretical approaches have been used to predict RMS matrix elements that are in qualitative agreement with experimental data, based on DDH
predictions of the coupling constants, and which seem to favour a small value of $h^1_\pi$ consistent with the $^{18}$F measurements shown in
Figure \ref{nuclearlimits}, but a comprehensive program to attempt to analyze all the data and set limits on the weak coupling constants has not
yet been attempted.

\subsection{The End of the Road}

Figures \ref{nuclearlimits} and \ref{couplings} summarize the present state of our knowledge using the meson-exchange framework. We have at
present reasonably clear experimental constraints  on four linear combinations of the six weak meson-nucleon couplings from a mixture of few
nucleon and finite nuclear experiments,  not all of which are in agreement with each other.  It is fair to say that the phenomenological
implications that one can draw from these results are unclear at best. On the one hand, the ${\vec p}p$ experiments now yield a rather stringent
set of constraints on the combinations of the DDH couplings $h_\rho^{pp}$ and $h_\omega^{pp}$ that govern the asymmetry over a fairly broad
range of energy. As we discuss below, $h_\rho^{pp}$ and $h_\omega^{pp}$ essentially parameterize  the contributions to $A_z$ from the
lowest-order, short-range PV potential in the EFT. To the extent that the energy-dependence of $A_z$ is dominated by that of the strong
interaction phase shifts that enter matrix elements of these operators as well as the unpolarized cross section, one need not think of
$h_\rho^{pp}$ and $h_\omega^{pp}$ as being specific to the DDH meson-exchange framework.

On the other hand, the nuclear PV experiments are largely sensitive to two different combinations of the DDH parameters shown in Figure
\ref{nuclearlimits}: $h_S^{\rm nuc}\equiv -(\hrhoz+0.7\homz)$ and $h_V^{\rm nuc}\equiv \hpi-0.12\hrhoo-0.18\homo$ \cite{Haxton:2001zq}. In the
past, it has been the conventional practice to project the constraints from the $A_z$ measurements onto the $h_{S,V}^{\rm nuc}$ plane by using
the DDH theoretical ranges for the $h_\rho^{1,2}$ and $\homo$. Here, however, we choose not to do so since we want to minimize the number of
theoretical assumptions used in the extraction of information from experiment. Instead, we treat the nuclear experiments separately from the
${\vec p}p$ measurements. In the ideal situation, the analysis of the nuclear  experiments would yield a self-consistent region for the
$h_{S,V}^{\rm nuc}$ -- a situation that clearly does not emerge from Figure \ref{nuclearlimits}. The primary problem seems to be the inclusion
of the $^{133}$Cs anapole moment constraint, which finds no region of simultaneous consistency with all the other nuclear PV experiments. To
explain this discrepancy, one might naturally re-examine the shell model  calculation leading to the cesium band. However,  theoretical
considerations suggest that a more realistic calculation would lead to an even larger discrepancy \cite{Haxton:2001mi,Haxton:2001zq}, while the
results of more naive shell model computations
\cite{Flambaum:1980sb,Flambaum:1984fb,Bouchiat:1990bu,Bouchiat:1991hw,Dmitriev:1994ct,Dmitriev:1999xq,Wilburn:1998xq,Auerbach:1999jb} lead to a
similar result. Evidently, additional insight into the many-body physics of nuclear PV is needed before a consistent phenomenology can be
obtained with the meson-exchange framework.

Even if such a consistent picture had emerged from experiment, extraction of fundamental information on the $\Delta S=0$ HWI would still be
problematic. To explain why, we consider the physics embodied by the $h_{S,V}^{\rm nuc}$. As compared with $h_\rho^{pp}$ and $h_\omega^{pp}$,
these effective nuclear couplings correspond to different combinations of the short range EFT operators and long-range $\pi$-exchange PV
potential than those that enter the ${\vec p}p$ asymmetry. However, the nuclear matrix elements of these operators sample the spatial-dependence
of both the operators as well as their action on the nuclear wavefunctions, and we have no simple way of disentangling the two as we do for the
${\vec p}p$ asymmetries.

To illustrate, we first consider the momentum space form of the $\rho$- and $\omega$-exchange operators appearing in $V^{\rm PV}_{DDH}$. Each
term contains a pseudoscalar of the general form ${\vec\sigma}_i\cdot {\vec p_j}$ -- where the subscripts refer to nucleons $i$ or $j$ -- times
a function of the momentum transfer $q=|{\vec q}|$ generated by the meson propagator in the static limit ($q_0=0$) and a hadronic form factor
$F_{\rho,\ \omega}(q^2)$ arising from the meson-nucleon vertices: \be \label{eq:vexp1} V^{\rm PV}_{V\ \rm -exchange}\ \sim \
{\vec\sigma_i}\cdot{\vec p}_j\ F_V(q^2)\ [1+q^2/m_V^2]^{-1} \ \ \  . \ee For $q << m_V$ we may expand in powers of $q^2/m_V^2$: \be
\label{eq:vexp2} V^{\rm PV}_{V\ \rm -exchange}\ \sim \ {\vec\sigma_i}\cdot{\vec p}_j\ F_V(0)\ \left(1+ \frac{q^2}{m_V^2}\left[m_V^2
F^\prime(0)-1\right] +\cdots \right) \ \ \  , \ee where the $+\cdots$ indicate higher order terms. In a model-independent approach, the
coefficient $[m_V^2 F^\prime(0)-1]$ of the $q^2/m_V^2$ term in Eq. (\ref{eq:vexp2}) would be replaced by an a priori unknown coefficient whose
value would have to be taken from either experiment or a QCD computation. A similar statement holds for the higher order terms. At each order
$n$ in the expansion, one could also include additional pseudoscalar operators having a different structure than ${\vec\sigma}_i\cdot{\vec p}_j
(q^2/m_V^2)^n$. In effect, the meson-exchange framework imposes model-dependent relations between all of the higher-order operator coefficients
and those  of the lowest order terms -- relations that may or may not hold in the SM.

The impact of the higher-order operators on nuclear matrix elements depends on both the values of the operator coefficients as well as the
spatial dependence of the nuclear wavefunctions. Extraction of the $h_{S,V}^{\rm nuc}$ from experimental observables relies on {\em both} the
relationships between these operators implicitly assumed by the meson-exchange model as well as on nuclear model-space truncations and other
nuclear structure inputs that affect the wavefunctions employed. At present, we have no rigorous way to disentangle the impact of either on the
extracted PV couplings, and the constraints in Figure \ref{nuclearlimits} may reflect both artifacts of nuclear structure calculations as well
as assumed operator relations.  If one seeks to study the fundamental $\Delta S=0$ HWI experimentally, then one would like to avoid such an
implicit reliance on model-dependent assumptions and nuclear structure inputs. The theoretically cleanest way to do so is to exploit effective
field theory -- wherein operator relations are determined systematically from experiment -- and by studying hadronic PV in few-body systems, for
which ab initio theoretical computations are available.

\section{Effective Field Theory Framework}
\label{sec:EFT}

Effective field theories are ideally suited to situations where there exists a distinct hierarchy of scales. In the presence instance, several
scales are relevant: the weak scale $v=(\sqrt{2} G_F)^{1/2} = 246$ GeV; the hadronic scale $\Lambda_{\rm HAD} \approx 1$ GeV; the pion mass and
decay constant, $m_\pi\approx 140$ MeV and $F_\pi=93.2$ MeV, respectively; and the typical momentum $Q$ relevant to a parity-violating hadronic
or nuclear process. The distance at which the repulsive core of the strong NN potential becomes dominant, $r\lsim 0.4$ fm corresponds to a mass
scale $\gsim 500$ MeV which, for our purposes, we take to be of order $\Lambda_{\rm HAD}$. The size of hadronic matrix elements relevant to the
HWI is governed by the the ratio of $F_\pi^2$ to $v^2$ that is typically normalized to the quantity $g_\pi$ as in Eq. (\ref{eq:gpidef}).

The remaining scales can be used to construct an effective
Lagrangian out of nucleon and pion fields, where the operators are
organized according to powers of $Q/\Lambda$. For processes in
which $Q << m_\pi$, one should take $\Lambda=m_\pi$, treating the
pions as heavy and \lq\lq integrating them out" of the effective
theory. For $Q\gsim m_\pi$, the pion must be kept as an explicit
degree of freedom, and one should take $\Lambda=\Lambda_{\rm
HAD}$. Consequently, we consider two versions of the EFT
corresponding to these two different regimes for $Q$.

\subsection{The Pionless Effective Field Theory} \label{sec:pionless}

In the PV EFT without pions, the lowest-order pseudoscalar operators contain four nucleon fields and are ${\cal O}(Q)$, since they must
transform as ${\vec\sigma}_i\cdot{\vec p}_j$. At this order, there nominally exist ten different contact operators. As shown in
Reference~\cite{Zhu:2004vw}, the most general short-range (\lq\lq SR") potential has the coordinate-space form
\begin{eqnarray}\label{3}\nonumber
V_{1,\ \rm SR}^{\rm PV} ({\vec r}) &=& {2\over \Lambda^3} \left\{
\left[ C_1 + (C_2+C_4) \left({\tau_1 +\tau_2 \over 2}\right)_3 +
C_3 \tau_1 \cdot \tau_2 +{\cal I}_{ab} C_5 \tau_1^a
\tau_2^b\right]
\right.\\
\nonumber && \left. \qquad \left( {\vec \sigma}_1 -{\vec
\sigma}_2\right)\cdot \{-i\vec{\nabla},f_m(r)\}
\right.\\
\nonumber && \quad \left. +  \left[ {\tilde C}_1 + ({\tilde
C}_2+{\tilde C}_4)\left({\tau_1 +\tau_2\over 2}\right)_3 + {\tilde
C}_3 \tau_1 \cdot \tau_2 +{\cal I}_{ab} {\tilde C}_5\tau_1^a
\tau_2^b \right]
\right.\\
\nonumber && \left. \qquad i \left( {\vec \sigma}_1\times {\vec
\sigma}_2 \right) \cdot [-i\vec{\nabla},f_m(r)]
\right.\\
\nonumber && \quad \left. + \left( C_2 -C_4 \right)
\left({\tau_1-\tau_2\over 2}\right)_3 \left( {\vec \sigma}_1
+{\vec \sigma}_2\right) \cdot \{-i\vec{\nabla},f_m(r)\}
\right.\\
&& \quad \left. + C_6 i\epsilon^{ab3} \tau_1^a \tau_2^b \left(
{\vec \sigma}_1 +{\vec \sigma}_2\right)\cdot
[-i\vec{\nabla},f_m(r)] \right\}
%\quad
\label{eq:vpvsr}
\end{eqnarray}
where ${\cal I}_{ab}={\rm diag}(1,1,2)$ and where the subscript
\lq\lq 1" on $V_{1,\ \rm SR}^{\rm PV} ({\vec r})$ essentially
indicates that this potential appears at ${\cal O}(Q^1)$ in the
EFT. In arriving at Eq. (\ref{eq:vpvsr}) we have introduced the
function $f_m({\vec r})$ that is strongly peaked about $r=0$ with
some width $\sim 1/m$ and goes to $\delta^{(3)}({\vec r})$ in the
zero-width  ($m\rightarrow\infty$) limit. For practical purposes,
we will take $1/m\lsim 0.4$ fm.

At first glance, one sees a dependence on ten a priori unknown constants $C_{1-6}$ and ${\tilde C}_{1-5}$ that encode information about the
short-distance weak interaction between two nucleons\footnote{The combination ${\tilde C}_2-{\tilde C}_4$ does not appear, so that only ten
combinations of the  eleven constants $C_{1-6}$ and ${\tilde C}_{1-5}$ appear in Eq. (\ref{eq:vpvsr}).}. When considering processes with $Q <<
m_\pi$, however, not all of the operators in Eq. (\ref{eq:vpvsr}) are independent. In this regime, PV observables are dominated by mixing
between S- and P-waves, for which there exist only five independent spin-isospin amplitudes:
\begin{itemize}
\item[i)] $d_t(k)$, representing ${}^{3}S_1(I=0)-{}^{1}P_1(I=0)$
mixing \item[ii)] $d_s^{0,1,2}(k)$, representing
${}^{1}S_0(I=1)-{}^{3}P_0(I=1)$ mixing generated by  $I=0,1,2$
operators respectively; and \item[iii)] $c_t(k)$, representing
${}^{3}S_1(I=0)-{}^{3}P_1(I=1)$ mixing
\end{itemize}
where we have used the notation of Danilov \cite{Danilov:1965pl} and Desplanques and Missimer \cite{Desplanques:1976mt, Desplanques:1998ak}. At
the low energies relevant to the pionless EFT, the energy dependence of these
 amplitudes is dominated by the strong interaction phase shifts.
 Denoting the spin singlet and spin triplet strong interaction S-wave
 scattering amplitudes as $m_s(k)$ and $m_t(k)$, respectively, we have
\bea \nonumber
d_t(k) & =& \lambda_t m_t(k) +\cdots \\
\label{eq:strongweak}
d_s^i(k) & = & \lambda_s^i m_s(k) +\cdots \\
\nonumber c_t(k) & =& \rho_t m_t(k) +\cdots \eea where the $+\cdots$ indicate small corrections to the energy dependence arising from the strong
P-wave phase shifts. In his early formulation of the problem, Danilov \cite{Danilov:1965pl} argued for the forms in Eq. (\ref{eq:strongweak}),
omitting the small corrections indicated. In the EFT framework, it is straightforward to derive the proportionality of the S-P amplitudes and
the $m_i(k)$ by computing the relevant T-matrix elements and summing up the strong rescattering contributions. In doing so, one finds that at
low energies where P-wave rescattering contributions are small, the $\lambda_i$ and $\rho_t$ are given by the ratio of the lowest order S-P
scattering amplitude to the lowest order, parity-conserving S-wave amplitude in a given spin-isospin channel \cite{Zhu:2004vw}.

The coefficients $\lambda_t$, $\lambda_s^i$, and $\rho_t$ are themselves proportional to various combinations of the $C_i$ and ${\tilde C}_i$
appearing  in $V_{1,\ \rm SR}^{\rm PV}({\vec r})$. In the zero-range ($m\rightarrow\infty$) limit, one has \footnote{ The last equation of Eqs.
(46) in Reference~\cite{Zhu:2004vw} contains an error. The sign in front of $C_6$ should be \lq\lq $+$" rather than \lq\lq $-$".}
\cite{Zhu:2004vw}
\begin{eqnarray}
\lambda_t & \propto & (C_1-3C_3) -({\tilde C_1}-3{\tilde C_3}) \nonumber \\
\lambda_s^0 & \propto & (C_1+C_3) +({\tilde C_1}+{\tilde C_3})\nonumber \\
\label{eq:lincomb}
\lambda_s^1 & \propto & (C_2+C_4) +({\tilde C_2}+{\tilde C_4}) \\
\lambda_s^2 & \propto & -\sqrt{8/3}(C_5 +{\tilde C_5})\nonumber\\
\rho_t & \propto & \frac{1}{2}(C_2-C_4) +C_6 \ \ \ .\nonumber
\end{eqnarray}
In effect, at low energies, five of the operators in Eq. (\ref{eq:vpvsr}) become redundant, leaving only five independent S-P amplitudes.

Inclusion of finite-range effects  leads to modifications of these relations. Arriving at exact relations using state-of-the-art NN potentials
remains an unfinished task for many-body theorists. However, we  provide approximate expressions by drawing on the work of Desplanques and
Benayoun \cite{Desplanques:1986cq}. We obtain
\begin{eqnarray}
m_N\rho_t & = & -\frac{2}{{\bar\Lambda}^3}\left[B_2\left(\frac{1}{2}C_2-\frac{1}{2}C_4+C_6\right)+ B_3\left(\frac{1}{2}C_2-\frac{1}{2}C_4-C_6\right)\right] \nonumber \\
m_N\lambda_t & = & -\frac{2}{{\bar\Lambda}^3}\left[B_4\left(C_1-3C_3+{\tilde C}_1-3{\tilde C}_3\right)
+B_5\left(C_1-3C_3-{\tilde C}_1+3{\tilde C}_3\right)\right] \nonumber \\
\label{eq:cirelations}
m_N\lambda_s^0 & = & -\frac{2}{{\bar\Lambda}^3}\left[B_6\left(C_1+C_3+{\tilde C}_1+{\tilde C}_3\right)
+B_7\left(C_1+C_3-{\tilde C}_1-{\tilde C}_3\right)\right]  \\
m_N\lambda_s^1 & = & -\frac{2}{{\bar\Lambda}^3}\left[B_6\left(C_2+C_4+{\tilde C}_2+{\tilde C}_4\right)
+B_7\left(C_2+C_4-{\tilde C}_2-{\tilde C}_4\right)\right] \nonumber \\
m_N\lambda_s^2 & = & \frac{4\sqrt{6}}{{\bar\Lambda}^3}\left[B_6\left(C_5+{\tilde C}_5\right) +B_7\left(C_5-{\tilde C}_5\right)\right] \nonumber
\end{eqnarray}
where ${\bar\Lambda}=m_N m_\rho^2/\Lambda^3$ and where the $B_k$ are linear combinations of the $\beta_{ij}^\pm$ of
Reference~\cite{Desplanques:1986cq}. For example, using the values of those constants obtained with the Reid Soft Core (RSC) potential, we
obtain $B_k=(-0.0043, 0.0005,-0.0009,-0.0022,-0.0067, 0.0003)$ for $k=2,\ldots, 7$, respectively\footnote{We emphasize that the expressions in
Eq. (\ref{eq:cirelations}) are applicable to the EFT without pions. For a discussion of the modifications due to inclusion of explicit pions,
see Sec. \ref{sec:eftpions} below. In particular, the lowest-order contribution from single pion exchange appears in $\rho_t$.}.

The coefficients  $B_{2,5,6}$ multiply the combinations of $C_i$ and ${\tilde C}_i$ that one expects to arise in the lowest order EFT [ Eq.
(\ref{eq:lincomb})]. The remaining terms are generated by finite-range contributions that occur beyond leading order. Although we have not
included the full set of PV operators and amplitudes that occur at next-to-leading order (NLO), the magnitudes of the $B_{3,4,7}$ -- relative to
the $B_{2,5,6}$ -- give an indication of the magnitude of higher-order effects and of the error associated with working to lowest order. The RSC
values obtained in Reference~\cite{Desplanques:1986cq} give $|B_3/B_2| = 0.12$, $|B_4/B_5|=0.41$, and $|B_7/B_6|=0.04$, suggesting that the
impact of neglected higher-order contributions are generally small except in the case of $\lambda_t$.

Before considering the application of this framework to specific observables, it is useful to obtain theoretical predictions for the quantities
$\rho_t$ and $\lambda_{s,t}$. To that end, we first consider the correspondence with the DDH meson-exchange model and delineate the relationship
between the $C_i$ and ${\tilde C}_i$  and the DDH parameters. Using \be f_m({\vec r})=\frac{m^2}{4\pi r}\ {\rm exp}(-mr) \ee with $m$ being the
parameter that defines the range of the PV potential, and letting \be \label{eq:ddheft0} {\bar \Lambda}_V^3\equiv \frac{\Lambda^3 }{ m_N m^2_M}
\ee for $M=\rho,\omega$ we have\footnote{The corresponding expressions given in Eq. (141) of Reference~\cite{Zhu:2004vw} contain typographical
errors. The quantity $\Lambda_V$ should contain only two powers of $m_M$ in the denominator and each of the $C_i$ should be proportional to the
product of a strong coupling $g_M$ and the relevant $h_M^i$ as in Eqs. (\ref{eq:ddheft0}-\ref{eq:ddheft2}) here.}
\begin{eqnarray}
\nonumber
 C_1^{DDH} =-\frac{1}{2} {\bar\Lambda}_\omega^3 g_\omega h_\omega^0\qquad &
 C_2^{DDH} =-\frac{1}{2} {\bar\Lambda}_\omega^3 g_\omega
h_\omega^1\\
\label{eq:ddheft1}
 C_3^{DDH} =-\frac{1}{2} {\bar\Lambda}_\rho^3 g_\rho h_\rho^0 \qquad &
C_4^{DDH} =-\frac{1}{2}{\bar\Lambda}_\rho^3 g_\rho h_\rho^1\\
\nonumber
 C_5^{DDH} =\frac{1}{4\sqrt{6}} {\bar\Lambda}_\rho^3 g_\rho
h_\rho^2\qquad & C_6^{DDH}= -\frac{1}{2} {\bar\Lambda}_\rho g_\rho h_\rho^{\prime 1}
\end{eqnarray}
and \bea \label{eq:ddheft2}
 {{\tilde C}_{i}^{DDH}\over C_{i}^{DDH}}&=&1+\chi_\omega\ \ \ i=1,2,\\
\nonumber
 {{\tilde C}_{i}^{DDH}\over C_{i}^{DDH}}&=&1+\chi_\rho\ \ \ i=3-5\ \ \ .
\eea
Using Eqs. (\ref{eq:ddheft1},\ref{eq:ddheft2}) one can easily
obtain the expressions for $\rho_t$, $\lambda_t$, and $\lambda_s^{0,1,2}$ and employ the DDH best values
and reasonable ranges  for the PV meson-nucleon couplings to obtain the predictions for the PV LECs
listed in Table  \ref{tab:PVLEC}. The results in columns 2-4 were obtained by retaining only the combinations of the $C_i$, ${\tilde C}_i$ that arise at lowest order in the EFT. Column five contains estimates of the size of higher-order contributions, based on the DDH best values for the $C_i$, ${\tilde C}_i$ and the the terms in Eqs.~(\ref{eq:cirelations}) proportional to $B_{3,4,7}$.

To obtain a sense of the possible variations in theoretical predictions for the PV LECs from their correspondence with the DDH parameterization,
we consider two approaches. First, we vary the values of the strong couplings in Eqs. (\ref{eq:ddheft1}) in accordance with the Bonn one pion
exchange potential  as suggested by Miller \cite{Miller:2003gn}. As indicated in Table \ref{tab:PVLEC}, doing so changes both the overall
magnitude of the $C_i$ and ${\tilde C}_i$ as well as the relation between the two and leads to generally wider ranges for the PV LECs than
obtained with the values of the strong couplings originally used by DDH. Second, we give expectations  using naive dimensional analysis (NDA)
considerations, as discussed in Section \ref{sec:eftpions} below. The NDA arguments suggest that the magnitudes of the $C_i$ and ${\tilde C}_i$
ought to be of order $16\pi^2\sim 150$, but do not fix the signs of the $C_i$ and ${\tilde  C}_i$. To translate the NDA estimates into
predictions for the five LECs, we simply take the magnitudes of the combinations of $C_i$ and ${\tilde C}_i$ appearing in Eqs.
(\ref{eq:cirelations}) to be the NDA expectation for any one of them.

\begin{table}[h]

\def~{\hphantom{0}}
\caption{{\em Predictions for the five, PV low energy constants (LECs) characterizing hadronic PV in the pionless EFT.  All values are quoted in
units of $g_\pi = 3.8 \times 10^{-8}$. Estimates are obtained using Reid Soft Core potential as in Reference \cite{Desplanques:1986cq} ;  DDH
values are taken from Reference \cite{Desplanques:1979hn}}} \label{tab:PVLEC}
\begin{center}
\begin{tabular}{@{}lccccc @{}}%
%\toprule
\hline \hline PV LEC & DDH  Best & DDH   range & DDH  plus Bonn & higher order & NDA
\\ \hline
%\colrule
$m_N\rho_t$        & 0.05      & 0.07 $\rightarrow$ 0.03  & 0.18 $\rightarrow$ 0.08    & $\pm 0.006$ & $\pm 0.38$     \\
$m_N\lambda_t$       & 0.84   & -1.00 $\rightarrow$ 2.48  & -2.44 $\rightarrow$ 6.12   &   $\pm 1.2$ & $\pm 0.31$  \\
$m_n\lambda_s^0$       & 3.82    & -4.86 $\rightarrow$  11.7 & -9.84 $\rightarrow$ 22.96    &   $\pm 0.09$  & $\pm 0.64$  \\
$m_N\lambda_s^1$       &  0.37   & 0.64 $\rightarrow$ 0.21   & 1.53 $\rightarrow$ 0.54   &  $\pm 0.0006$ & $\pm 0.64$  \\
$m_N\lambda_s^2$     & 2.72     & 2.17 $\rightarrow$ 3.15    & 3.83 $\rightarrow$ 5.55   &   $\pm 0.06$  & $\pm 3$   \\

%\botrule
\hline
\end{tabular}
\end{center}
\end{table}

\vspace*{-0.5cm}

We note that in obtaining the correspondence between the PV LEC's and the DDH predictions in the meson-exchange model, we have not included
contributions from the parameter $h_\rho^{1 \prime}$ that appears in $V^{\rm PV}_{\rm DDH}$. Using the estimate of
Reference~\cite{Holstein:1981cg} for this parameter would increase the magnitude of $\rho_t$ appearing in Table \ref{tab:PVLEC}, though not
substantially. Generally, analyses of hadronic PV using the meson-exchange model have neglected this parameter since its contribution to $V^{\rm
PV}_{\rm DDH}$ has the same spin-isospin structure as for the $\pi$ exchange contribution but is suppressed by its short range. In the EFT
framework, this term corresponds to the operator proportional to $C_6$ that -- along with $C_{2,4}$ -- contributes to $\rho_t$. Since the
coefficients of $C_{2,4,6}$ in $\rho_t$ have comparable magnitude, we see no model-independent reason to neglect the $C_6$ contribution in the
most general analysis.

With the foregoing set of benchmarks in hand, it is instructive to consider the dependence of various few-body PV observables on the $\lambda_i$
and $\rho_t$ and to outline a program of high precision measurements that could be used to determine these parameters. The few-body PV
observables of interest include:

\begin{itemize}

\item  Polarized $\vec{p} p$ scattering at 13.6 and 45 MeV, yielding the asymmetry $A_z^{pp}$,

\item  Polarized ${\vec p}\alpha$ scattering at 46 MeV, giving $A_z^{p\alpha}$,

\item  Radiative $\vec{n} p$ capture at low energy: ${\vec n}p\to d\gamma$, yielding the photon asymmetry $A_\gamma^d$,

\item Radiative $np$ capture with unpolarized neutrons, giving the photon circular polarization $P_\gamma^d$, or alternatively, the asymmetry
$A_L^\gamma$ in ${\vec\gamma} d\to np$,

\item Rotation through an angle $\phi$ about the momentum direction of polarized neutron spin passing through $^4$He, from which one extracts
the quantity $d\phi^{n\alpha}/dz$,

\item Radiative capture of polarized neutrons on deuterium at threshold, ${\vec n}d\to t\gamma$, yielding the photon asymmetry $A_\gamma^t$.

\end{itemize}

Explicit expressions for these quantities in terms of the S-P amplitude parameters have been given in references~\cite{Desplanques:1976mt,
Desplanques:1998ak} and elsewhere. In Table \ref{tab:pvobs}, we give the coefficients of the five PV LECs as they appear in  various
observables. Theoretical expectations for these observables in the pionless EFT can be obtained using Tables \ref{tab:PVLEC}   and
\ref{tab:pvobs}. We emphasize that these expectations will, in general, differ when going to the EFT with explicit pions discussed in Section
\ref{sec:eftpions}. In particular, the parameter $m_N\rho_t$ that governs the asymmetry $A_\gamma^d$ will be dominated by LO pion exchange,
assuming $h_\pi^1$ has its natural size.   In Table \ref{tab:pvobs} we also show the most precise experimental results that have been published
to date for these observables. Notably, only $A_z^{pp}$ and $A_z^{p\alpha}$ have been measured to sufficient precision to establish a nonzero PV
effect that can be used to constrain the PV LECs.  A review of recent, ongoing and prospective efforts to obtain precision measurements of these
important few-body PV effects appears in Section \ref{sec:horizons} below.\footnote{The measurement of $P_\gamma^d$ is particularly challenging
because of the limited sensitivity of conventional $\gamma$ ray circular polarimeters in the $\sim $ MeV energy range;  an alternative
measurement of $A_L^\gamma$ close to threshold yields the same physics and may be more accessible to experiment.  A possible experiment is at
the early stages of development in Athens  \cite{Stiliaris:2005ab}, as mentioned in Section \ref{sec:exptprog}, but is not discussed further in
Section \ref{sec:horizons}.}

\begin{table}[h]
\def~{\hphantom{0}}
\caption{{\em Sensitivities of selected PV observables to the five PV LECs.  The first column gives the observable, while subsequent columns
give the coefficients of a given PV LEC.  For the $pp$ asymmetry, $k$ is the incident proton momentum in the lab frame.  The final columns give
the most precise experimental limits to date and their references;  for the $pp$ case, we quote the 13.6 and 45 MeV measurements, to which the
lowest order EFT best applies. Note that the $p\alpha$ asymmetry is evaluated at 46 MeV; $d\phi^{n\alpha}/dz$ is evaluated in rad/m.
}}\label{tab:pvobs}

\begin{center}
\begin{tabular}{@{}c c c c c c  c  l @{}}%
%\toprule
\hline \hline Observable & $m_N\rho_t$ & $m_n\lambda_t$ & $m_N\lambda_s^0$ & $m_N\lambda_s^1$ & $m_N\lambda_s^2/\sqrt{6}$ & Expt. ($ 10^{-7}$) &
Ref.
\\ \hline
%\colrule
&&&&&& \\
$A_z^{pp}(k) $        & 0      & 0  &  $4k/m_N$    & $4k/m_N$ & $4k/m_N$  &  $-0.93 \pm 0.21$ &
~\cite{Eversheim:1991tg}\\
&&&&&& $-1.50 \pm 0.22$ &  ~\cite{Kistryn:1987tq} \\
&&&&&& \\
$A_z^{p\alpha}$       & -1.07   & -0.54  & -0.72   &  -0.48 & 0 &   $-3.3 \pm 0.9$  &  ~\cite{Lang:1985jv} \\
&&&&&& \\
$P_\gamma$        & 0   & 0.63  & -0.16   &  0 & 0.32 & $1.8 \pm 1.8$ &~\cite{knyazkov84} \\
&&&&&& \\
$A_\gamma^{d}$       & -0.107   & 0  & 0   &  0 & 0  &  $0.6 \pm 2.1 $ &~\cite{Cavaignac:1977uk}\\
&&&&&& \\
$d\phi^{n\alpha}/dz$      & -2.68   & 1.34  & 1.8   &  -1.2 & 0  & $8 \pm 14 $ & ~\cite{markoff97} \\
&&&&&& \\
$A_\gamma^{t}$       & -3.56   & -1.39  & -0.95   &  -0.24 & 1.18 &  $42 \pm 38 $ &~\cite{Alberi:1988fd}
\\
&&&&&& \\

%\botrule
\hline
\end{tabular}
\end{center}
\end{table}

It should be remarked that the longitudinal asymmetry $A_z^{pd}$ in ${\vec p}d$ scattering was measured to high precision at 43 MeV
\cite{Kloet:1983ta,Kistryn:1989ad} in the 1980's, but remains to be analyzed in a theoretical framework accounting for PV in both elastic
scattering and breakup channels, both of which contributed to the experimental signal. There exist also several additional possibilities for
few-body experiments, for which we have not yet obtained expressions for the PV observables in terms of the PV LECs. These include measurements
of PV neutron spin rotation on hydrogen, $d\phi^{np}/dz$  and deuterium, $d\phi^{n d}/dz$, and one could also consider the circular polarization
$P_\gamma^t$ in $nd \rightarrow t \gamma$ as well as perhaps even the transmission asymmetry of unpolarized neutrons through polarized $^3$He.
With realistic prospects for performing some or all of these measurements in the future, deriving the appropriate expressions would clearly be
important.\\

\noindent{\bf {\em Limits of the pionless EFT}}

Before discussing the EFT with pions, it is instructive to investigate the limits of validity of the pionless theory by considering the the
low-energy ${\vec p}p$ asymmetry as an illustrative example. Again, considering only S-P mixing, it is straightforward to show that
\cite{Zhu:2004vw}:
\begin{equation}
\label{eq:azlowe} A_z^{pp}={\sigma_+-\sigma_-\over \sigma_++\sigma_-} ={4k\, {\rm Re}[m_s^*(k)d_s^{pp}(k)] \over |m_s(k)|^2}\simeq
4k\lambda_s^{pp}\ \ \ .
\end{equation}
where where we have neglected small corrections arising from the P-wave phase shifts as before. A more complete decomposition of $A_z^{pp}$ in
terms of higher partial waves was first worked out by Simonius \cite{Simonius:1988rg} and subsequently studied by several authors. A recent
analysis using state-of-the-art NN potentials was performed in Reference~\cite{Carlson:2001ma}  and used to extract the DDH parameters
$h_\rho^{pp}$ and $h_\omega^{pp}$.

Following Simonius' original formulation, one may write the asymmetry as \be \label{eq:azpartial} A_z^{pp} = \sum_{{\rm even}\ J} f_{J^\pm}(E)
K_{J^\pm}(E,\theta) \ \ \ , \ee where the \lq\lq $\pm$" indicate the orbital angular momentum $L=J\pm 1$ and where $E$ and $\theta$ are the
energy and scattering angle, respectively.  The $f_{J^\pm}$ are \lq\lq reduced" PV transition amplitudes and the $K_{J^\pm}$ contain all the
dependence on the strong phases that arise from rescattering. For the $J=0$ partial wave, only $f_{0^+}$ exists, and it is proportional to the
combination $d_s^{pp}$ of the S-P amplitudes $d_s^i$ that enter the ${\vec p}p$ process: \be d_s^{pp} = -2i f_{0^+}\ \exp i\left[\delta(^1{\rm
S}_0)+\delta(^3{\rm P}_0)\right] \ \ \ . \ee To the extent that one may neglect the finite (but short) range of the PV potential, $d_s^{pp}$ and
$f_{0^+}$ are proportional to $\lambda_s^{pp}$. Moreover, in obtaining Eq. (\ref{eq:azlowe}), we have neglected the dependence of $d_s^{pp}$ on
$\delta(^3{\rm P}_0)$. Doing so is equivalent to taking $K(E,\theta)\propto k$.

While this approximation holds to a high degree for accuracy for low energy interactions, it breaks down for $E\sim 100$ MeV ($k\sim 300$ MeV).
Due to cancellations between the effects of the S- and P-wave phase shifts in $d_s^{pp}$ that occur above this energy, the contribution of the
$J=0$ partial wave to $A_z^{pp}$ falls rapidly, going to zero at $E=227$ MeV. Moreover, the contribution from the $J=2$ partial wave --
dominated by the $^{1}{\rm D}_2$--$^{3}{\rm P}_2$ ($f_{2^-}$) mixing -- becomes appreciable. Taking advantage of the former, the beam energy for
the TRIUMF $A_z^{pp}(221\ {\rm MeV})$ measurement was optimized, accounting for finite acceptance of the apparatus, to ensure that that the
$f_{0^+}$ contribution was entirely cancelled, so that the asymmetry was determined almost entirely by the $f_{2^-}$ contribution
\cite{Berdoz:2002sn, Carlson:2001ma}. Thus, a combined analysis of the TRIUMF and lower-energy $A_z^{pp}$ measurements yields constraints on the
two amplitudes $f_{0^+}$ and $f_{2^-}$. In effect, one may treat the values of $h_\rho^{pp}$ and $h_\omega^{pp}$ obtained in
Reference~\cite{Carlson:2001ma} as equivalent parameterizations of these two partial wave transition amplitudes and need not tie them
specifically to the meson-exchange framework.

Clearly, at  the energies of the recent TRIUMF experiment, the lowest-order EFT is no longer applicable, and one most include  ${\cal O}(Q^3)$
operators in $V_{\rm SM}^{\rm PV}$ that characterize P-D mixing. Doing so introduces a host of  new, a priori unknown operator coefficients. At
the same time, the redundancy of operators in the ${\cal O}(Q)$ $V_{\rm SR}^{\rm PV}$ in Eq. (\ref{eq:vpvsr}) that holds for the S-P amplitudes
breaks down, and all ten lowest order operators become independent. Determining all of these constants from experiment would be unrealistic, so
we will restrict our attention to the energy range where the lowest order EFT applies.

\subsection{PV EFT with Pions} \label{sec:eftpions}

For PV processes involving few-body nuclei, the relevant energy scale $Q$ is no longer set solely by experimental kinematics, but also includes
the relevant internal momentum of the bound nucleons. Since the latter can be as large as the Fermi momentum of $\gsim 200$ MeV, it is no longer
reasonable to treat the pion as heavy. For decades, phenomenological strong interaction potentials for light nuclei have included a long-range
$\pi$-exchange component. In the case of the EFT formulation, treating pionic contributions consistently has presented challenges. The
difficulty arises from the presence of two-nucleon poles in iterated $\pi$-exchange amplitudes whose contributions are enhanced by $\sim m_N/Q$
relative to naive expectations. These enhanced contributions spoil the \lq\lq power counting" in $Q/\Lambda$ that is essential to the success of
the EFT approach. Thus, one must sum strong pion exchange to all orders in order to obtain a consistent treatment. Indeed, as shown in
Reference~\cite{Fleming:1999ee}, treating the pion perturbatively as in the framework of references \cite{Kaplan:1998tg, Kaplan:1998we} does not
lead to a convergent expansion in all channels of the NN interaction.

An alternate formulation, originally proposed by Weinberg \cite{Weinberg:1990rz, Weinberg:1991um}, entails performing the all-orders resummation
in terms of the effective strong potential. In order to be self-consistent, however, it appears that one must simultaneously perform the EFT
expansion of the potential about the chiral limit \cite{Beane:2001bc, Beane:2002vs, Beane:2002xf}: $m_\pi=0$, since a full resummation of the
chiral symmetry-breaking component of the one pion-exchange (OPE) potential leads to inconsistent renormalization \cite{Kaplan:1998tg,
Kaplan:1998we, Beane:2001bc, Beane:2002vs, Beane:2002xf}. Although the chiral expansion treatment of the Weinberg approach is still under
development, we follow Reference~\cite{Zhu:2004vw} and employ it here.

The basis for the EFT with pions is chiral perturbation theory ($\chi$PT) whose formalism is well-known and will not be repeated here. However,
we note that its formulation for hadronic PV processes was first written down by Kaplan and Savage \cite{Kaplan:1992vj}, whose notation follow.
The non-linear dependence of the effective Lagrangian on the pion is implemented via the field
\begin{equation}
\xi  = \exp \left( i\pi^a \tau^a\over 2 F_\pi  \right),
\end{equation}
where $\pi^a$, $a=1, 2, 3$ are the isospin components of the pion
field. Defining the quantity
\begin{equation}
\label{eq:xminus} X_{-}^3 = \xi^{\dag}\tau^3\xi - \xi\tau^3\xi^{\dag} \;\; ,
\end{equation}
the lowest-order Lagrangian for a PV interaction of the pion with
a single nucleon is
\begin{eqnarray}
\label{eq:pvyuk} {\cal L}^{(-1)}_{\pi N,PV}&=&-{\hpi\over
2\sqrt{2} }{\bar N}
X_{-}^3 N\nonumber\\
&=&-i\hpi (\bar p n \pi^+ -\bar n p \pi^-) +\cdots
\end{eqnarray}
where the \lq\lq $+\cdots$" indicate the higher order terms in odd
powers of $(\pi^a \tau^a/F_\pi)$ that arise from expanding the
exponential in $\xi$. The leading term in the  Lagrangian in Eq. (\ref{eq:pvyuk}) is
identical to the PV Yukawa interaction in the DDH model.
Consequently, its contribution to the PV NN potential will be the
same as the first term in Eq. (\ref{eq:DDH1}):
\be
\label{eq:vpvlr-1} V^{\rm PV}_{(-1,\ {\rm LR})}(\vec{r}) = i{\hpi
g_{A}m_N\over \sqrt{2}F_\pi}\left({{\vec\tau_1}\times{\vec\tau_2}\over
2}\right)_3 (\vec{\sigma}_1+\vec{\sigma}_2)\cdot \left[{{\vec
p}_1-{\vec p}_2\over 2m_N},w_\pi (r)\right]\
\ee
The \lq\lq $-1$"
subscript indicates that this long-range (\lq\lq LR") potential is
${\cal O}(Q^{-1})$, a feature most readily seen from its
momentum-space form:
\begin{equation}
\label{eq:vpvlr-1mom} V^{PV}_{(-1, {\rm LR})} ({\vec q})= -i{g_A
\hpi\over \sqrt{2} F_\pi} \left({{\vec\tau}_1\times {\vec \tau}_2\over 2}\right)_3 {
\left( {\vec \sigma}_1 +{\vec \sigma}_2\right)\cdot {\vec q} \over
q^2 +m_\pi^2}\ \ \ ,
\end{equation}
where ${\vec q}={\vec p}_1-{\vec p}_1^{\ \prime} ={\vec p}_2^{\ \prime}-{\vec p}_2$ is the three-momentum of the exchanged pion, ${\vec p}_i$
(${\vec p}_i^{\ \prime}$) is the initial (final) momentum of nucleon $i$, and $q=|{\vec q}|$.

Since the PV potential must transform as a pseudoscalar, the operators in it will contain odd numbers of derivatives. Thus, one would expect the
sub-leading terms to be ${\cal O}(Q)$, as in $V^{\rm PV}_{(1, \ {\rm SR})}$. In principle, loop corrections to the Lagrangian (\ref{eq:pvyuk})
or to $V^{\rm PV}_{(-1,\ {\rm LR})}(\vec{r})$ could bring in a factor of $m_\pi$ or $p^2/m_\pi$, leading to an ${\cal O}(Q^0)$ component.
Explicit computations, however, indicate that no such contributions exist. Consequently, the subleading components of the potential start off at
${\cal O}(Q)$ (next-to-next-to-leading order, NNLO), and it is convenient to distinguish them according to their range:

\begin{itemize}

\item[i)] {\bf Short range:} $V^{\rm PV}_{(1, {\rm SR})}$ as in Eq. (\ref{eq:vpvsr}) but with $\Lambda=\Lambda_{\rm HAD} \approx
\Lambda_\chi=4\pi F_\pi$.

\item[ii)] {\bf Medium range: } $V^{\rm PV}_{(1, \ {\rm MR})}$, generated by the two pion-exchange diagrams of Figure \ref{fig:mexvseft}c) and
proportional to $\hpi$. The structure of the operator is most conveniently given in momentum space, as it carries a non-analytic dependence on
pion momentum and mass \cite{Zhu:2004vw}:
\begin{eqnarray}\label{v3}\nonumber
V_{(1,\ \rm  MR)}^{PV} ({\vec q})  &=& -{1\over \Lambda_\chi^3}
\left\{
  {\tilde C}^{2\pi}_2(q)  {\tau_1^z +\tau_2^z\over 2}
i \left( {\vec \sigma}_1 \times {\vec \sigma}_2 \right) \cdot
{\vec q} \right.
\nonumber \\
\label{eq:vpvmr}
&& \quad \left. + C^{2\pi}_6(q) i\epsilon^{ab3} [{\vec\tau_1}
\times{\vec\tau_2}]_3 \left( {\vec \sigma}_1 +{\vec
\sigma}_2\right)\cdot {\vec q} \right\},
\end{eqnarray}
where
\begin{eqnarray}\label{good}\nonumber
& {\tilde C}^{2\pi}_2(q) =4\sqrt{2} \pi g_A^3 \hpi L(q) \\
\label{eq:tpecoeff}
& C^{2\pi}_6 (q)= -\sqrt{2}\pi g_A \hpi L(q) + {3\sqrt{2}\over
2}\pi \left[ 3L(q) -H (q)\right] g_A^3 \hpi,
\end{eqnarray}
and
\begin{eqnarray}\label{LandH}
L(q) &=& {\sqrt{4m_\pi^2 +q^2} \over q} \ln \left(
{\sqrt{4m_\pi^2 + q^2}+ q \over 2m_\pi}\right),
\nonumber\\\
H(q) &=& {4m_\pi^2 \over 4m_\pi^2 +q^2} L(q).
\end{eqnarray}

\item[iii)]{\bf  Long range:} $V^{\rm PV}_{(1, \ {\rm LR})}$, generated by one-loop corrections to the PV $\pi NN$ Yukawa and parity conserving
strong vertices as well as by additional operators having a distinct structure from the lowest order potential. As discussed in
Reference~\cite{Zhu:2004vw}, the impact of all but one of the NNLO PV  operators can be absorbed in to $V^{\rm PV}_{(-1, \ {\rm LR})}$ and
$V^{\rm PV}_{(1, \ {\rm SR})}$ through a suitable redefintion of the operator coefficients. The remaining PV operator and PC operators give rise
to the momentum space potential \cite{Zhu:2004vw}
\begin{eqnarray}
\nonumber
 V^{\rm PV}_{1,\ \rm LR}({\vec p}_1,
\cdots, {\vec p}_2^{\ \prime})   &=& { g_{A}k_{\pi }^{1a}\over
\Lambda_\chi F_\pi^2}\left({{\vec \tau}_1\times {\vec\tau}_2\over
2}\right)_3\Biggl[
 {{\vec\sigma}_1\cdot{\vec p}_1^{\ \prime}\times{\vec p}_1 {\vec\sigma}_2\cdot
{\vec q}_1\over q^2+m_\pi^2}+(1\leftrightarrow
2)\Biggr] \\
\label{eq:vpvlrnnlomom}
&& +i{g_A
\hpi\over \sqrt{2} m_N^2 F_\pi} \left({{\vec\tau}_1\times {\vec \tau}_2\over 2}\right)_3 {1\over q^2 +m_\pi^2}
\Biggl\{ \\
\nonumber
&& \frac{1}{4}\left[(|{\vec p}_1|^2-|{\vec p}_1^{\ \prime}|^2){\vec \sigma}_1\cdot({\vec p}_1^{\ \prime}+{\vec p}_1)-(1\leftrightarrow 2)\right] \\
\nonumber
&&-\frac{1}{8}\left[(|{\vec p}_1|^2+|{\vec p}_1^{\ \prime}|^2){\vec \sigma}_1\cdot{\vec q}+(1\leftrightarrow 2)\right]\\
\nonumber
&&+\frac{1}{4}\left[{\vec\sigma}_1\cdot{\vec p}_1^{\ \prime}\ {\vec q}\cdot{\vec p}_1+
{\vec\sigma}_1\cdot{\vec p}_1\ {\vec q}\cdot{\vec p}_1^{\ \prime}+(1\leftrightarrow 2)\right]\Biggr\}
\end{eqnarray}
where  ${\vec q}_i={\vec p}_i^{\ \prime}
-{\vec p}_i$ and $k_{\pi }^{1a}$ is a constant that must be determined from experiment.

\end{itemize}

In addition to considering the potential through ${\cal O}(Q)$, one must also include two--body current operators that contribute to the same
order when considering PV processes involving photons. These operators include the standard PV meson-exchange currents associated with the
$\pi$-exchange potential and those arising from the covariant derivatives in  $V^{\rm PV}_{(1,\ {\rm SR})}$. In addition, there exists a new,
independent current operator \cite{Zhu:2004vw}
\begin{equation}\label{eq:mecnew}
{\vec J}({\vec x}_1, {\vec x}_2, {\vec q}) = {\sqrt{2} g_A {\bar
C}_\pi m_\pi^2\over \Lambda^2 F_\pi} {\rm e}^{-i{\vec
q}\cdot{\vec x}_1}\ \tau_1^+\tau_2^-\ {\vec\sigma}_1\times {\vec
q} \ {\vec\sigma}_2\cdot{\hat r} \ H_\pi(r) \ + \
(1\leftrightarrow 2),
\end{equation}
where
\begin{equation}
H_\pi(r) = {\exp{(-m_\pi r)}\over m_\pi r} \left(1+{1\over m_\pi
r}\right),
\end{equation}
and ${\bar C}_\pi$ is an additional LEC parameterizing the leading
PV $NN\pi\gamma$ interaction.

\vskip 0.2in

\noindent{\bf \em  PV EFT with pions: new features}

In comparison with the DDH meson-exchange potential, the EFT with explicit pions introduces several qualitatively distinct features. First, the
${\cal O}(Q)$, long range operators in Eq. (\ref{eq:vpvlrnnlomom},\ref{eq:mecnew}) have no analog in the DDH framework and introduce two new
unknown constants, $k_\pi^{1a}$ and ${\bar C}_\pi$. Given the novel nature of these operators, the impact of their contributions to the
observables discussed above has yet to be determined with explicit, few-body computations. Based on the isospin structure of these two
operators, however, one would expect them to contribute to the same processes, such as ${\vec n}p\to d\gamma$, that are sensitive to the LO
$\pi$-exchange potential. Power counting implies that the magnitude of their contribution should be smaller than that of $V^{\rm PV}_{(-1, \
{\rm LR})}$, but their long range character suggests that they should have a greater impact than the operators in $V^{\rm PV}_{(1,\  {\rm
SR})}$. Future few-body calculations should test these expectations.

The two $\pi$-exchange (TPE) medium range potential, $V^{\rm PV}_{(1, \ {\rm MR})}$, is similarly a new feature of the EFT framework. In the
past, others have attempted to introduce PV TPE using model frameworks (see, {\em e.g.}, Reference~\cite{Pirner:1973wy}), but to our knowledge,
the result in Eq.~(\ref{eq:vpvmr}) gives the first formulation that is model-independent and consistent with the symmetries of QCD. The operator
coefficients ${\tilde C}^{2\pi}_2$ and $C^{2\pi}_6$ have been labelled to indicate their correspondence with the operators in $V^{\rm PV}_{(1,
{\rm SR})}$, but as indicated in Eq.~(\ref{eq:tpecoeff}), the coefficients are fixed in terms of $\hpi$ and are not independent free parameters.
The operator proportional to $C^{2\pi}_6$ has the same isospin structure as $V^{\rm PV}_{(-1, {\rm LR})}$ and will contribute to any process --
such as ${\vec n}p\to d\gamma$ -- that is sensitive to the LO $\pi$-exchange potential. The operator proportional to ${\tilde C}^{2\pi}_2$ has
the same structure as the $h_\omega^1 (1+\chi_\omega)$ term in the DDH potential and will, therefore, generate a medium range contribution to
any observable sensitive to the latter combination in the meson-exchange model. In particular,   this operator will contribute to both the
$f_{0^+}$ and $f_{2^-}$ partial wave terms in $A_z^{pp}$, implying that this observable is sensitive to $\hpi$ at the same order in $Q$ as the
the short-range effects enter. In either case, proper inclusion of TPE medium range potential will affect the determination of $\hpi$ obtained
from a global analysis of few-body PV observables.

 In principle, one should also take into account three-body PV forces
 that arise in the EFT with pions, as we will discuss few-body experiments in systems
 involving three or more nucleons. As discussed in Reference \cite{Zhu:2004vw},
 three-body PV forces do not arise at ${\cal O}(Q)$, so we restrict our attention to the two-body PV EFT interaction.\\

\noindent{\bf \em  PV LECs: naive dimensional analysis}

Beyond these qualitative observations, we are not able to make any quantitative statements regarding the relative importance of the new features
of the PV EFT with pions. Evaluating their contributions to specific observables now constitutes an open problem for few-body theorists. That
being said, it is instructive to estimate the size of the constants $C_i$, ${\tilde C}_i$, $\hpi$, $k_\pi^{1a}$, and ${\bar C}_\pi$ that arise
in the EFT. A systematic way of doing so -- known as \lq\lq naive dimensional analysis" (NDA) -- was developed by Georgi and
Manohar~\cite{Manohar:1983md} and has successfully explained the size of a variety of LECs in $\chi$PT. According to NDA, one should construct
operators by scaling fields and derivatives to their natural scales:
\begin{equation}
\label{eq:georgi}
\left({D_\mu \over\Lambda_\chi}\right)^d
\left({\pi\over F_\pi}\right)^p
\left({{\bar N} N\over\Lambda_\chi F_\pi^2}\right)^{f/2}
\times (\Lambda_\chi F_\pi)^2
\times (g_\pi)^n \ ,
\end{equation}
where $d$, $p$, $f=2k$, $k$ and $n$ are positive integers and where $g_\pi$ is as given in
Eq. (\ref{eq:gpidef}). Consequently, one expects the PV LEC's to have the magnitudes
\begin{eqnarray}
\label{hpiNDA}
\hpi &\sim& \left(\frac{\Lambda_\chi}{F_\pi}\right) g_\pi\sim 10 g_\pi\\
\label{CiNDA}
C_i, {\tilde C}_i &\sim&
\left(\frac{\Lambda_\chi}{F_\pi}\right)^2  g_\pi\sim 100 g_\pi\\
\label{CbarNDA}
k_{\pi}^{1a},\ {\bar C}_\pi & = & g_\pi\ \ \sim g_\pi \ \ \  .
\end{eqnarray}
It is interesting that since $\Lambda_\chi/F_\pi=4\pi\sim 12$, the NDA estimates for
$\hpi$ and the $C_i$, ${\tilde C}_i$ are roughly equal to the expectations based on
correspondence with the DDH best values. In contrast, one expects the new long range
LECs $k_\pi^{1a}$ and ${\bar C}_\pi$ to be an order of magnitude smaller than $\hpi$.
This difference, however, simply reflects the conventions used above in normalizing
the various operators.

\vskip 0.2in

\noindent{\bf \em PV EFT: phenomenology}

The phenomenology of the PV EFT with pions is clearly more challenging than for the pionless theory, since one encounters additional unknown
parameters when working to ${\cal O}(Q)$, and since a variety of new contributions remain to be computed. In principle, there exists a viable
experimental program that could determine the constants to this order, including the six measurements highlighted in Section \ref{sec:pionless}
for the pionless theory plus two additional, independent experiments. As noted in Section \ref{sec:pionless}, possibilities for the latter
include a careful analysis of the ${\vec p}d$ scattering asymmetry results for $A_z^{pd}$ \cite{Kloet:1983ta,Kistryn:1989ad} and future neutron
spin rotation measurements on hydrogen or deuterium. Recently, the possibility of performing PV photo- and electro-production experiments on the
single nucleon to determine the of $\pi N$ and $\gamma N$ couplings has received considerable attention. Carrying out these experiments -- which
we discuss below -- would provide additional, independent input for the determination of the PV LECs.

Theoretically, the extraction of these constants from experiment will require new calculations to determine the contributions from (a) the
medium-range, TPE potential (\ref{eq:vpvmr}), (b) the NNLO single pion-exchange potential (\ref{eq:vpvlrnnlomom}), and (c) the meson-exchange
current (\ref{eq:mecnew}). At present, only the dependence on $h_\pi^1$ generated by LO pion-exchange is known: \be \label{eq:rhot} m_N\rho_t =
1.04 h_\pi^1 + m_N \rho_t^{\rm SR} + m_N \Delta\rho_t \; , \ee where $\rho_t^{\rm SR}$ gives the dependence of the $^3S_1$-$^3P_1$ mixing on the
constants appearing in $V^{\rm PV}_{1,\ {\rm SR}}$ as in Eq. (\ref{eq:cirelations}); $\Delta\rho_t$ gives the presently unknown contributions
generated by $V^{\rm PV}_{1,\ {\rm MR}}$ and $V^{\rm PV}_{1,\ {\rm LR}}$; and the  first term on the right side of Eq. (\ref{eq:rhot}) is
generated by $V^{\rm PV}_{-1,\ {\rm LR}}$ as computed by Desplanques and Benayoun using the Reid Soft Core potential \cite{Desplanques:1986cq}.
Since $V^{\rm PV}_{1,\ {\rm MR}}$ contains spin-isospin structures corresponding to both the $C_6$ and $C_2$ terms in $V^{\rm PV}_{1,\ {\rm
SR}}$, the LEC $\lambda_s^1$ will also contain a dependence on $h_\pi^1$ generated by two-pion exchange. In addition to computing these new
contributions to $\rho_t$ and $\lambda_s^1$, theorists must also determine new meson-exchange current contributions to processes such as ${\vec
n}+p\to d+\gamma$,  generated associated with $V^{\rm PV}_{1,\ {\rm MR}}$ and $V^{\rm PV}_{1,\ {\rm LR}}$ and required by gauge invariance, as
well as the contribution from the new current in Eq. (\ref{eq:mecnew}).

Assuming that a successful  program is completed and the complete set of PV LECs through ${\cal O}(Q)$ are extracted from experiment, the values
of these parameters would the provide model-independent benchmarks for Standard Model theory. In this case, the theoretical challenge would be
analogous to the one encountered with $\chi$PT for pseudoscalar mesons, where for example, at ${\cal O}(Q^4)$, there exist ten independent LECs
that have been determined from experiment. The theoretical task is now to explain how the dynamics of QCD give rise to the values of these
constants, and to that end, a number of approaches have been pursued. Ultimately, of course, one would like to compute these constants using
lattice QCD, but given the difficulties in putting two or more hadrons on the lattice,   approaches based on symmetry arguments or models are an
attractive, interim alternative. A particularly fruitful direction involves taking the limit of QCD with a large number of colors ($N_C$),
wherein one expects the exchange of heavy mesons, such as the $\rho$ and $\omega$, to dominate the underlying QCD dynamics of the LECs. In the
$N_C=3$ world that we inhabit, this large-$N_C$ picture of \lq\lq resonance saturation" works remarkably well in accounting for the values of
the constants. It remains to be understood why large the large $N_C$ limit is so successful in this case, and future lattice QCD computations
should address this problem\footnote{At the same time, it has become important to know the values of certain ${\cal O}(Q^6)$ constants that
presently cannot be taken from experiment and that are needed for the extraction of the Cabibbo-Kobayashi-Maskawa matrix element $V_{us}$ from
$K_{e3}$ decay data. The insights and techniques developed to explain the ${\cal O}(Q^4)$ constants will be essential in obtaining reliable
theoretical values for the unknown higher order terms.}.

In the case of the PV LEC's, the results from experiment should teach us whether the large $N_C$ resonance saturation picture applies to the
$\Delta S=0$ HWI involving baryons as well as to strong interactions between light mesons. In effect, the DDH model {\em assumes} the validity
of resonance saturation, albeit with a truncated spectrum of exchanged mesons that may or may not reflect accurately the underlying dynamics. As
discussed at the outset of this article, there exists ample evidence that QCD symmetry arguments fall short when confronting the phenomenology
of the $\Delta S=1$ HWI, so one should apply caution when adopting another one (viz, large $N_C$) to predict weak, hadronic $\Delta S=0$
processes. At the same time, one would like to derive as much model-independent information as possible on the HWI in each sector, so that one
can gain new insights into the puzzles associated with strangeness-changing processes. For example, if one ultimately found  a set of PV LECs
that agreed with expectations based on NDA and resonance saturation, one might conclude that the breakdown of symmetry-based expectations in the
$\Delta S=1$ sector is associated with the dynamics of the participating strange quark. On the other hand, should the PV LECs depart
substantially from NDA and large $N_C$ expectations, one would look elsewhere to determine dynamics general to all sectors of the HWI.

\subsection{Recent Theoretical Work}

As the foregoing discussion makes evident, there now exists ample motivation
for new theoretical work within the context of the EFT for hadronic PV. The past decade
has seen initial efforts in this direction, and we review some of this work here.

\vskip 0.2in

\noindent{\bf \em Few-Body Systems.}

In the two-body sector, recent interest has focused on the asymmetry $A_\gamma^d$ for ${\vec n}p\to d\gamma$, where new computations using EFT
and Green's function Monte Carlo methods have been used. The lowest order EFT computation yields the asymmetry \cite{Kaplan:1998xi} \be
\label{eq:npdgeft1} A_\gamma^d =  -\frac{2 m_N}{\gamma^2}\ \frac{{\rm Re}[(X+Y)^\ast W]}{2|X|^2+|Y|^2} \ee where $X$ and $Y$ give contributions
to the parity-conserving ${\vec n}p\to d\gamma$ amplitude for an initial $^3 S_1$ and $^1 S_0$ state, respectively, and $W$ gives the PV
amplitude: \be \label{eq:npdgeft2} W=-g_A\hpi \frac{\sqrt{\pi\gamma}}{2\pi F_\pi}\ \left[\frac{m_\pi}{(m_\pi+\gamma)^2}
-\frac{m_\pi^2}{2\gamma^3}\ln\left(\frac{2\gamma}{m_\pi}+1\right)+\frac{m_\pi^2}{\gamma^2(m_\pi+\gamma)}\right]\ \ \ , \ee with
$\gamma=\sqrt{m_N B}$ and $B$ being the deuteron binding energy \footnote{In writing down Eq.~(\ref{eq:npdgeft2}), we used an overall sign that
is opposite the one appearing in Reference ~\cite{Kaplan:1998xi}, thereby following the conventions used elsewhere in the literature
\cite{Desplanques:2000ej,Savage:2000iv}.}. From these expressions, one obtains $A_\gamma^d=-0.17\hpi$. The coefficient of $\hpi$ in this result
is nearly a factor of two larger than in previous, wavefunction-based computations, and stimulated considerable follow-up theoretical activity
\cite{Desplanques:2000ej,Hyun:2001yg,Schiavilla:2002uc,Liu:2002bq,Schiavilla:2004wn}. In particular, the authors of Reference~\cite{Hyun:2001yg}
computed the asymmetry to NLO in the Weinberg scheme, using two-body wavefunctions derived from the Argonne $v_{18}$ potential, and obtained
$A_\gamma^d=-0.10\hpi$, in close agreement with previous results obtained using Siegert's theorem, $A_\gamma^d\simeq -0.11\hpi$. Subsequently,
the authors of Reference~\cite{Schiavilla:2002uc} performed a wavefunction-based computation in the DDH framework, using Argonne $v_{18}$,
Nijmegen-I, and Bonn-CD wavefunctions and found $A_\gamma^d=-( 0.106 \to 0.109) \hpi +\cdots$, where the range corresponds to the choice of
different potentials and where the $+\cdots$ indicate small, ${\cal O}(10^{-9})$ contributions from the short-range terms in the DDH potential.

From a more academic perspective, several computations of the deuteron anapole moment have been performed using EFT and wavefunction methods
\cite{Savage:1998rx,Savage:1999cm,Khriplovich:2000mb,Hyun:2002in,Liu:2003au}. The LO EFT result is \cite{Savage:1998rx}: \be F_A^{(D)} =-\frac{e
g_A \hpi m_N^2}{24 F_\pi}\ \left[\kappa_1\frac{m_\pi+\gamma}{(m_\pi+2\gamma)^2}+\frac{2m_\pi+9\gamma}{6(m_\pi+2\gamma)^2}\right] \ee where
$\kappa_1\simeq 1.85$ is the isovector anomalous magnetic moment of the deuteron. The numerical value of $F_A^{(D)}$ obtained from this
expression has the same sign but a magnitude that is $\sim 40-50\%$ larger than results obtained using a wavefunction computation
\cite{Khriplovich:2000mb} or Weinberg EFT approach \cite{Hyun:2002in}. From a practical standpoint, the impact of the deuteron anapole moment
would be most relevant to the interpretation of PV elastic ${\vec e}d$ scattering, wherein it would generate a potentially important
contribution to the isoscalar axial vector response \cite{Musolf:1993tb}. Since the latter vanishes at tree-level in the Standard Model, it is
particularly transparent to higher-order effects, such as electroweak radiative corrections, the strange quark axial vector current, and
hadronic PV. To date, however, no experiments have been proposed to study the PV elastic deuterium asymmetry.

\vskip 0.2in

\noindent{\bf\em Single Nucleon Sector.}

From the standpoint of theoretical interpretability, PV pion photo- and electroproduction processes involving single nucleon targets offer
several advantages. In particular, they provide a means for accessing the PV $\pi NN$ couplings directly without having to disentangle the
short- and medium-range effects discussed above. Moreover, the use of $\chi$PT to describe low-energy pion-nucleon interactions is well
established and has been thoroughly studied in the parity conserving sector. Looking to future work in QCD, it is likely that attempts to
compute the PV $\pi NN$ couplings on the lattice will precede any efforts to study the short range PV interaction.

Historically, single nucleon, PV pion photo- and electroproduction processes were studied in the meson-exchange framework two decades ago by
Woloshyn \cite{Woloshyn:1978qk} and Li and Henley \cite{Li:1982fd}. These authors found that one should expect PV asymmetries for the scattering
of longitudinally polarized photons to be of order a few $\times 10^{-7}$ assuming the DDH best value for $\hpi$, while  those for
electroproduction could be up to two orders of magnitude larger. Given the experimental challenges associated with such tiny asymmetries, and
the prospect of measuring considerably larger effects in light nuclei, the prospects for carrying out single nucleon studies were largely
ignored for many years.

Recently, however, advances in experimental techniques for measuring ${\cal O}(10^{-7})$ photo- and electroproduction asymmetries have
stimulated renewed interest in this direction. Theoretically, Chen and Ji reformulated the earlier work for near threshold PV  pion photo-
\cite{Chen:2000hb} and electroproduction \cite{Chen:2000km} using heavy baryon $\chi$PT. Subleading contributions were subsequently considered
by the authors of Reference~\cite{Zhu:2000hf}.  To ${\cal O}(Q)$, one has for the threshold photoproduction asymmetry \be
\label{eq:thresholdphoto} B_\gamma=\frac{\sqrt{2}F_\pi}{g_A m_N}\left[\mu_p-\mu_n\left(1+\frac{m_\pi}{m_N}\right)\right] \hpi
+\frac{4\sqrt{2} m_\pi}{g_A\Lambda_\chi} {\bar C}_\pi \\
\ee where $\Lambda_\chi=4\pi F_\pi$ and the terms proportional to $m_\pi$ give the ${\cal O}(Q)$ contributions. A measurement at Jefferson
Laboratory could in principle yield a result for $B_\gamma$ with statistical accuracy at better than the $10^{-7}$ level;  however, substantial
technical challenges associated with the implied current mode pion detection would have to be overcome to design a successful experiment.

Considerably larger photo- and electroproduction asymmetries may be observed in the vicinity of the $\Delta(1232)$ resonance. The asymmetry in
this region is dominated by the lowest order PV $\gamma N\Delta$ interaction that does not contribute strongly at lower energies
\cite{Zhu:2001br}: \be \label{eq:ddelta} {\cal L}_{\rm PV}^{\Delta N\gamma} = i\frac{e}{\Lambda_\chi}\left [d_\Delta^+ {\bar\Delta}_\mu^+
\gamma_\lambda p+ d_\Delta^- {\bar \Delta}_\mu^0 \gamma_\lambda n\right] F^{\mu\lambda} +{\rm h.c.} \ee where the $d_\Delta^\pm$ are low energy
constants that govern the strength of the PV transition. For $E_\gamma\approx m_\Delta-m_N$, the PV photoproduction asymmetry is \be
B_\gamma^\pm \approx -\frac{2 d_\Delta^\pm}{C_3^V}\frac{m_N}{\Lambda_\chi} +\cdots \ee where $C_3^V\sim 2$ is the transition magnetic moment and
where the \lq\lq $+\cdots$" indicate higher order, chiral corrections. The latter have been computed in Reference~\cite{Zhu:2001re} and shown to
be relatively small. Thus, the size of the resonance asymmetry is essentially set by $d_\Delta^\pm$.

The authors of Reference~\cite{Zhu:2001re} noted that a determination of $d_\Delta^\pm$ could provide additional insights into the puzzles
surrounding the $\Delta S=1$ HWI discussed earlier. In particular, if the dynamics responsible for the enhanced PV hyperon radiative decay
asymmetries also occur for the $\Delta S=0$ PV $N\to \Delta$ transition, then one might expect $B_\gamma^\pm$ as large as a few $\times
10^{-6}$. Such an effect could be observed in forward angle PV electroproduction experiments, for which the contribution of $Z^0$ exchange
becomes kinematically suppressed, thereby exposing the $d_\Delta$ contribution \cite{Zhu:2001re}. On the other hand, if the occurrence of
enhanced PV asymmetries is unique to the $\Delta S=1$ sector and is associated with presence of valence strange quarks, then one would expect
$B_\gamma^\pm$ to smaller by an order of magnitude or more.

With this motivation in mind, possibilities for measuring $B_\gamma^\pm$ using both the G0 and $Q_{weak}$ instrumentation at Jefferson Lab are
being actively pursued.  The G0 collaboration will measure inclusive pion asymmetries in an upcoming backward angle run on a deuterium target
\cite{Martin03}.  A measurement of the PV asymmetry in inclusive inelastic $ep$ scattering  at much lower $Q^2$  to $\simeq 0.09$ ppm is also
envisioned as a future enhancement of the $Q_{weak}$ experimental program \cite{LOI-Tony}.

As an alternative to photo- and electroproduction, one may also consider PV Compton scattering from the nucleon. Computations of the asymmetry
for scattering with either polarized protons or polarized photons have been carried out in references ~\cite{Bedaque:1999dh} and
\cite{Chen:2000mb}. The asymmetry in both cases is proportional to $\hpi$ at leading order. For $E_\gamma << m_\pi$ and center of mass
scattering angle $\theta=\pi/2$ one has \cite{Chen:2000mb} \be A_{{\vec\gamma}p\to\gamma p}\sim 8.8\times 10^{-9}\left(\frac{\hpi}{5\times
10^{-7}}\right)\left(\frac{E_\gamma}{ 70\ {\rm MeV}}\right)^3 \ee where one expects higher order corrections to yield corrections of order
$25\%$ and where $\hpi$ has been scaled to the magnitude expected from NDA. Thus, one could expect to see an asymmetry of order a few $\times
10^{-8}$ -- roughly the size of the asymmetry expected in ${\vec n}+p\to d+\gamma$. The magnitude of the asymmetry for Compton scattering with a
polarized target is similar. To date, no experimental proposals have been developed for measuring either asymmetry.

\vskip 0.2in

\noindent{\bf\em Computing $\hpi$ in QCD}

Much of the recent theoretical focus has fallen on extracting the leading PV pion-nucleon coupling in a way that does not require knowledge of
the other PV LEC's or of many-body nuclear physics. The interest in $\hpi$ has also stimulated new analyses of QCD predictions for this quantity
that attempt to go beyond the work of DDH. A first principles computation will ultimately require use of lattice QCD, and while we are not aware
of immediate plans to carry out such a calculation, some of the necessary theoretical groundwork has been recently laid. In particular,
tractable lattice computations typically involve use of quarks that are heavier than the physical light quarks, so in order to obtain a
physically realistic QCD prediction from a lattice result, one needs to know the quark mass dependence of a given quantity\footnote{The advent
of chiral quarks has reduced the range over which an extrapolation must be performed.}. To that end, $\chi$PT provides the necessary link, since
the chiral expansion in powers of $m_\pi$ is equivalent to an expansion in $\sqrt{m_q}$.

\begin{figure}[hbt]
\centerline{\epsfig{figure=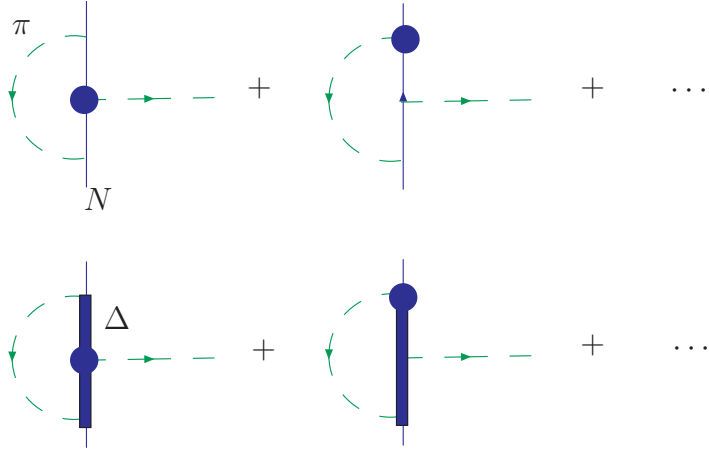,width=4.75in}} \caption{\em { One loop contributions to renormalized, PV $\pi NN$ Yukawa coupling.
 }}
\label{fig:chiral}
\end{figure}

The first such analysis of $\hpi$ was performed by the authors of Reference~\cite{Zhu:2000fc}, who employed SU(2)$_L\times$SU(2)$_R \;\; \chi$PT
with explicit $\Delta$ isobar degrees of freedom and computed all contributions to  ${\cal O}(Q^3)$. These contributions arise from one-loop
diagrams of the type illustrated in Figure \ref{fig:chiral}. Naively, one expects the loop contributions to be suppressed by powers of
$(Q/\Lambda_\chi)^k$ for $k=2,3$, where $\Lambda_\chi=4\pi F_\pi\sim$ 1 GeV and $Q$ is either $m_\pi$ or $m_\Delta-m_N$. In the case of $\hpi$,
however, the one-loop contributions receive logarithmic and fortuitous numerical enhancements, leading to the renormalized coupling \be
\label{eq:hpichiral} \hpi = 0.5 \mathring{h}^1_\pi + 0.25 h_A^1-0.24 h_\Delta +0.08 h_A^\Delta\ \ \ , \ee where $\mathring{h}^1_\pi$ is the bare
$\pi NN$ PV Yukawa coupling, $h_\Delta$ is the analogous PV $\pi N\Delta$ Yukawa coupling, $h_A^1$ parameterizes the isovector $\pi NN$ PV
derivative coupling \be {\cal L}_A^{\pi NN} = i\frac{h_A^1}{F_\pi^2} {\bar N}\gamma^\mu\gamma_5 N \left(\pi^+ D_\mu \pi^- -\pi^-
D_\mu\pi^+\right)+\cdots \ee and $h_A^\Delta$ parameterizes analogous PV $\pi N\Delta$ derivative couplings.

The corrections appearing in Eq.~(\ref{eq:hpichiral}) are only those having non-analytic quark mass dependence $m_q\ln m_q$ or $m_q^{3/2}$ (in
the $m_\Delta=m_N$ limit) and are uniquely identified with chiral loops\footnote{Terms that are analytic in $m_q$ can be absorbed into
corresponding terms in the Lagrangian}. As discussed in Reference~\cite{Zhu:2000fc}, terms of this form cannot arise in quark model matrix
elements of the hadronic weak Hamiltonian that were used in the analysis of DDH. Moreover, the \lq\lq sum rule" contribution to $\hpi$ that DDH
derived from $\Delta S=1$ decays using SU(6)$_w$ symmetry relied on tree-level symmetry relations that do not contain $m_q$-dependent symmetry
breaking effects generated by chiral loops. Thus, it appears unlikely that DDH benchmark estimates for $\hpi$ fully reflect the impact of its
quark mass dependence. Interestingly, the magnitudes of the coefficients of the various terms in Eq.~(\ref{eq:hpichiral}) are comparable,
allowing for possible cancellations between them that could reduce the magnitude of $\hpi$ from the DDH \lq\lq best value". In principle, a
study of $\hpi$ on the lattice could allow one to identify the unknown constants appearing in Eq.~(\ref{eq:hpichiral}) by varying both $m_q$ and
the number of colors.

At present, carrying out such an analysis with unquenched QCD is prohibitively expensive. An alternative approach that allows one to extrapolate
lattice computations to the domain of the physical, light quarks is to vary the valence and sea quark masses independently -- a technique known
as partial quenching. In order to identify the $m_q^{\rm valence}$ and $m_q^{\rm sea}$ dependence analytically from one-loop computations, one
must generalize standard $\chi$PT to the corresponding partially-quenched effective theory \cite{Sharpe:2000bc}. A computation using this
framework has been performed in Reference~\cite{Beane:2002ca}, leading to an analogous expression to that of Eq.~(\ref{eq:hpichiral}) that gives
the non-analytic valence and sea quark mass dependence in the partially-quenched theory. An extrapolation to the chiral domain also requires
inclusion of analytic terms that can be obtained from the effective Lagrangian and that were not written down explicitly in
Reference~\cite{Beane:2002ca}.

As of this writing, the first-principles QCD analysis of $\hpi$ has not advanced beyond the analytic work of references
\cite{Zhu:2000fc,Beane:2002ca}. Given the new experimental efforts, the time is clearly right for an investment in a lattice computation of
$\hpi$. On a longer term horizon, one would also hope to see lattice predictions of the constants $C_i$ and ${\tilde C}_i$ that parameterize the
${\cal O}(Q)$ short-distance, PV four-nucleon operators discussed above. Obtaining such computations will likely depend on progress in lattice
calculations of the low-energy, strong NN interaction. A program aimed in this direction is being carried out by Savage and collaborators (M.J.
Savage, private communication).

In the absence of first principles QCD computations of the PV constants, one may look to nucleon model calculations for guidance as to their
magnitudes. The expectations derived from the DDH  SU(6)$_w$/quark model treatment \cite{Desplanques:1979hn} has been discussed above. Alternate
approaches have recently been used to predict $h_\pi^1$. In the three-flavor Skyrme model of Reference \cite{Meissner:1998pu}, the dominant
contribution arises from terms in the $\Delta S=0$, PV four-quark Hamiltonian that contain strange quark bilinears such as ${\bar s}\gamma^\mu s
{\bar u}\gamma_\mu\gamma_5 u$. These terms sample the \lq\lq kaon cloud" that arises from the Wess-Zumino action via rotations of the chiral
soliton in SU(3) space. The resulting prediction is $2 g_\pi \lsim h_\pi^1 \lsim 3.4 g_\pi$. The corresponding prediction in the two-flavor
Skyrme model is considerably smaller. A prediction for a somewhat larger value has been obtained using QCD sum rules \cite{Henley:1995ad},
wherein one expands nucleon correlators in a pion background in terms of various quark and gluon condensates. Values for the latter -- such as
$\langle {\bar q}i\tau^a\gamma_5 q\rangle_\pi$ -- are taken from other nucleon properties, such as the strong $\pi NN$ coupling, leading to
$h_\pi^1\sim 8 g_\pi$. Both sets of model predictions are roughly consistent with expectations based on NDA as well as the updated  \lq\lq best
values" obtained from the SU(6)$_w$/quark model approach \cite{Desplanques:1979hn, Feldman:1991tj}

\subsection{Experimental Prospects}
\label{sec:horizons}

In the foregoing sections, we have emphasized the need for a complete set of precise PV measurements in the NN and few nucleon systems that can
be cleanly interpreted in terms of constraints on PV LECs.  As summarized in Table~\ref{tab:pvobs}, we have currently two significant
measurements in hand carried out with low energy proton beams ($A_z^{pp}$ and $A_z^{p\alpha}$), and two currently underway at existing neutron
facilities ($A_\gamma^d$ and $d \phi^{n\alpha}/dz$).  As noted earlier, there is also an existing $pd$ asymmetry measurement
\cite{Kistryn:1989ad} which poses additional theoretical challenges accounting for the angular dependence in both elastic scattering and breakup
channels, that remains to be analyzed in a common framework.

Some immediate prospects for improvement in neutron beam measurements will take advantage of the superior features of a high intensity pulsed
beam facility, the Spallation Neutron Source (SNS), currently under construction at Oak Ridge, Tennessee.  The SNS is anticipated to provide the
world's most intense beams of cold pulsed neutrons, approaching or even surpassing the time averaged intensities of cw reactor sources
\cite{Greene05} by 2008. A crucial advantage of pulsed beams over reactor sources for precision PV experiments results from the introduction of
new diagnostic capabilities via time-of-flight analysis of the neutron energy -- especially important for reducing $\gamma$-ray backgrounds and
systematic error diagnosis as well as neutron polarization diagnostics. In addition, the construction of a new facility allows for the
incorporation of technological advances in neutron guide instrumentation, particularly the use of high efficiency bent supermirror transport
guides which eliminate direct line of sight between the apparatus and the cold moderator, thereby reducing $\gamma$-ray and neutron backgrounds
without significant loss of beam flux. The SNS is building a new dedicated beamline for Fundamental Neutron Physics (FnPB) \cite{Greene05,
Huffman05} which is optimized to the needs of a
suite of precision experiments in HWI and neutron beta decay.\\

\noindent {\bf Neutron Capture Gamma Asymmetry Measurements} \\
As discussed earlier, the LANSCE phase of the ongoing NPDGamma experiment to measure $A_\gamma^{d}$ will be statistics limited at the 10$^{-7}$
level, while the expected asymmetry based on the NDA and DDH `best value' estimates for  $h_\pi^1$ is $A_\gamma^{d} = -5 \times 10^{-8}$.
NPDGamma is expected to be a key element of the initial SNS fundamental neutron physics program \cite{NPDGSNS}.  At the time of writing, the
apparatus has been commissioned, and systematic errors have been extensively studied at LANSCE, awaiting installation of the liquid parahydrogen
target. A conclusion of these studies is that the apparatus is ready to make a measurement of $A_\gamma^{d}$ with a statistical error of $1
\times 10^{-8}$. Systematic errors are expected to be at or below the $10^{-9}$ level;  however, additional running time to measure the
potential false asymmetry from aluminum vacuum windows is needed to reach the ultimate experimental precision that can be achieved at the SNS.

The SNS FnPB beamline has been extensively studied via Monte Carlo simulations, and reasonably conservative flux estimates indicate that a
measurement of $A_\gamma^{d}$ at the $1 \times 10^{-8}$ level should be possible in approximately 5000 hours of running on the new beamline once
the SNS reaches 1.4 MW operation. The experiment can be moved with only minimal changes required to the apparatus, since the beam conditions
(apart from increased flux) will be quite similar to those at LANSCE.  It is planned to use the NPDGamma apparatus to commission the new FnPB
beamline when it comes on line, beginning in 2008.  In addition to the higher beam flux, two significant improvements in experimental conditions
at the SNS are anticipated that could further improve the experimental precision.  Gamma background reduction will be achieved in part as a
result of the SNS curved neutron guide, and in part with the incorporation of additional lead shielding upstream of the main detector array.
Higher beam polarization should also be possible, via a combination of increased laser pumping power and/or the use of spectrally narrowed
lasers for optical pumping of the $^3$He spin filter cell, which have been demonstrated to produce up to 75\% polarization in bench tests as
compared to 40-55\% $^3$He polarization routinely achieved during extended running in FP12 at LANSCE \cite{NPDGSNS}.

As noted earlier, the PV gamma asymmetry $A_\gamma^{t}$ in the reaction $\vec{n} + d \rightarrow t + \gamma$ provides a complementary window on
the $\Delta S = 0$ HWI to its counterpart in the $np$ system.  A Letter of Intent has been submitted to the SNS to develop this experiment as a
logical follow up to the NPDGamma experiment \cite{NDTGSNS}.  As indicated in Table~\ref{tab:pvobs}, $A_\gamma^{t}$ displays a much larger
sensitivity to the PV LEC's -- including a $\sim 30$ times stronger sensitivity to $h_\pi^1$  -- than does  $A_\gamma^{d}$. Based on the various
estimates for the PV LECs (Table~\ref{tab:PVLEC}), one could expect $A_\gamma^{t}$ to be one to two orders of magnitude larger than
$A_\gamma^{d}$.

In principle, the $nd$ asymmetry measurement can be performed using most of the components of the NPDGamma apparatus, with an obvious exception
of the target.  The experiment is technically much more challenging than the $np$ case due to the much smaller $nd$ capture cross section;
consequently, most of the neutrons will scatter out of the target rather than being captured to produce the gamma rays of interest. A room
temperature liquid D$_2$O target is under consideration, with a length optimized between two competing factors -- the very small $nd$ capture
cross section, and the desire to avoid significant neutron depolarization in the target.  A novel target vessel and shielding scheme will be
required to absorb the scattered neutrons without producing significant background gamma rays.  The interactions of polarized cold neutrons in
D$_2$O are at present not well understood, and a program of detailed simulations and test measurements will be carried out in order to optimize
the design of the experiment.  Other target possibilities include cold solid orthodeuterium or solid ortho-D$_2$O in an effort to minimize
depolarization by slowing down the neutron beam, thereby increasing the capture probability (W.~M.~Snow, private communication).   An earlier
measurement of $A_\gamma^t$ was carried out at ILL and found a result consistent with zero: $A_\gamma^t = (4.2 \pm 3.8) \times 10^{-6}$
\cite{Avenier86, Alberi88}; the goal of the proposed SNS measurement is to reach a sensitivity $4 \times 10^{-7}$. Many of the systematic
effects are similar to those for the $np$ experiment, where they have been extensively studied.  It should be noted that the absolute tolerance
for systematic errors is relaxed for the $nd$ experiment, since the absolute precision goal is
more than an order of magnitude less stringent than that for the $np$ experiment.\\

\noindent{\bf Neutron Spin Rotation Measurements}\\ As for the neutron capture asymmetry measurements, the higher neutron flux anticipated for
the SNS as well as the pulsed nature of the beam offer compelling advantages for improving the precision of the $n\alpha$ PV spin rotation
measurement,  as well as the possibility of carrying out experiments on $np$ and perhaps even $nd$ spin rotation. To date, a Letter of Intent
has been submitted for the $n\alpha$ experiment \cite{nalphaloi}, anticipating a measurement accuracy of $1 \times 10^{-7}$ rad/m in 12 months
of data taking, which would represent a factor of three improvement over what is currently expected at NIST, with improved systematic error
control.   It is likely that the $n\alpha$ spin rotation experiment will run at the SNS  quite early in the FnPB experimental program.  The
collaboration intends in the longer term to pursue $np$ spin rotation at the SNS as well \cite{Markoff05}.  The sensitivity of the $n\alpha$
spin rotation experiment to the five PV LEC's in the pionless EFT is given in table \ref{tab:pvobs} ( predictions in the DDH meson-exchange
model are available for the $np$ and $n\alpha$ cases, as mentioned in Section \ref{sec:nspinrot}).

The basic spin rotation experimental technique has been outlined in Section \ref{sec:nspinrot}; recall the crucial role of the `$\pi$ coil' in
Figure \ref{spinrot}, which is used to suppress the effect of the much larger spin precessions due to residual magnetic fields inside the
apparatus. Since previous experiments have been carried out at reactor facilities with polychromatic cold neutron beams, the best one could do
was to optimize the field in the $\pi$ coil to rotate the neutron spins on average by 180$^\circ$ between the upstream and downstream segments
of the apparatus. Nearly a factor of two improvement can be obtained at a pulsed source \cite{Markoff05}  by ramping the $\pi$ coil current as a
function of time of flight to provide a 180$^\circ$ spin rotation for all neutron velocities.  In addition, one can take advantage of the fact
that the PV spin rotation is independent of neutron velocity, whereas scattering and magnetic field effects are in general energy and thus
velocity dependent for diagnosis of systematic effects.  For the hydrogen and possible deuterium target measurements, neutron depolarization
must be avoided by preparing cryogenic liquid targets in spin-selected molecular states - paramolecular for hydrogen and orthomolecular for
deuterium - an additional complication, particularly considering the requirement for constant cycling of the liquid between ``front'' and
``rear'' target locations. For neutron energies below 15 meV, depolarization does not occur in parahydrogen.  Less is known about the case for
orthodeuterium, but a  recent measurement at PSI found negligible depolarization of cold neutrons on orthodeuterium for a 4 cm target
(W.~M.~Snow, private communication), which is encouraging news for the prospect of a deuterium spin rotation measurement.

\section{Beyond Hadronic Parity Violation}

Throughout this article, we have emphasized the significance of hadronic PV as a probe of the strangeness conserving HWI. Here, we comment
briefly on its implications for other processes. We have already encountered one such process -- PV electron scattering -- discussed in
Section~\ref{subsec:mecmodel}. There, contributions from hadronic PV to the PV asymmetries constitutes a theoretical background that one must
compute reliably in order to extract information on other quantities of interest, such as the strange quark form factors.

A second illustration concerns neutrinoless double $\beta$-decay ($0\nu\beta\beta$). This process has considerable current interest, as it
provides the only known way to determine whether or not neutrinos are Majorana fermions (for recent reviews, see {\em e.g.}, references
~\cite{McKeown:2004yq,Elliott:2004hr}.) In addition, one also hopes to use $0\nu\beta\beta$ to determine the absolute scale of neutrino mass,
complementing what we know about $m_\nu$ from tritium $\beta$-decay and about neutrino mass differences from oscillation studies.

Unfortunately, the latter use of $0\nu\beta\beta$ is complicated by possible contributions to the rate from the exchange of heavy Majorana
particles, such as a heavy Majorana neutrino or the neutralinos of supersymmetry. In order to determine the absolute scale of $m_\nu$, one must
know the amplitude $A_H$ for heavy particle-exchange contributions, which can be comparable in magnitude to the amplitude $A_L$ for light
Majorana neutrino-exchange. From simple dimensional arguments, one has \cite{Cirigliano:2004tc} \be \label{eq:zeronuratio} \frac{A_H}{A_L} \sim
\frac{M_W^4 {\bar k}^2}{\Lambda^5 m_{\beta\beta}}\ \ \ , \ee where $m_{\beta\beta}$ is the effective mass of the light Majorana neutrino having
typical virtuality  ${\bar k}^2\sim (50\ {\rm MeV})^2$ and $\Lambda$ is the mass scale associated with the heavy Majorana particles. Given what
we know about $\Delta m_\nu^2$ and the neutrino mixing matrix elements that help determine $m_{\beta\beta}$, the ratio in
Eq.~(\ref{eq:zeronuratio}) can be ${\cal O}(1)$ for $\Lambda$ of order 1 TeV. Thus, it is important to analyze the possible heavy particle
contributions with theoretical clarity.

Recently, an EFT approach for doing so was developed by the authors of Reference~\cite{Prezeau:2003xn}. At the lepton-quark level, the effective
operators for heavy, Majorana particle exchange factorize into products of four-quark and two-lepton operators. The four-quark operators are
analogous to those entering the $\Delta S=0$ HWI, differing only in their respective representations in chiral SU(2). Consequently, the mapping
of these operators onto effective, hadronic operators involving nucleon and pion degrees of freedom is similar to the one used in obtaining the
EFT for hadronic PV.  Thus, the study of hadronic PV should provide insights into the EFT for heavy particle contributions in $0\nu\beta\beta$.
We consider two aspects of this correspondence in particular:

\begin{itemize}

\item[(i)] In contrast to the situation for hadronic PV, there does not exist a program
of few-body experiments from which one can determine the operator coefficients for the
$0\nu\beta\beta$ EFT. These coefficients must be computed theoretically. Comparisons of
analogous computations of the LECs for hadronic PV with experimental values should provide
new guidance for obtaining reliable computations in the $0\nu\beta\beta$ case. Indeed, as
argued above, hadronic PV provides a unique tool for learning how the strong interaction
dresses four-quark weak interactions into hadronic operators and amplitudes.

\item[(ii)] Again in contrast to hadronic PV, the only systems in which $0\nu\beta\beta$
can occur are heavy, complex nuclei. While there has been considerable recent progress in
understanding how an EFT power-counting of operators translates into a power-counting of
few-body matrix elements, the situation with complex nuclei is less clear. In order for
the EFT to provide realistic guidance to the size of heavy particle contributions to
$0\nu\beta\beta$ transition matrix elements, we need to know how well operator power counting
works for nuclear matrix elements.

In principle, hadronic PV can provide useful insights into this issue. Since the lowest order PV interaction can be determined from a program of
few-body experiments as outlined above, this interaction can then be used as a known probe of nuclear PV observables, such as the nuclear
anapole moment or the PV $\gamma$-decays of p-shell nuclei. To the extent that the lowest order PV interaction suffices to yield successful
descriptions of these nuclear PV observables, we would have evidence for the applicability of the EFT to nuclei. To the extent that it does not,
we would conclude that many-body renormalization of the lowest-order weak effective interaction is substantial and that high-momentum components
of nuclear wavefunctions play a more important role than one might naively expect. Either way, the implications for $0\nu\beta\beta$ would be
important.

\end{itemize}

\section{Summary and Conclusions}

The quest to explain the manifestations of weak interactions between quarks in strongly-interacting systems remains an important piece of \lq\lq
unfinished business" for Standard Model physics. In both the $\Delta S=1$ and $\Delta S=0$ sectors , the non-perturbative character of
low-energy QCD has been the stumbling block. The $\Delta S=1$ decays of hyperons in particular seem to elude explanation using the standard
symmetry and  effective field theory approaches that have been so successful in treating low-energy strong interactions. Whether these puzzles
simply reflect the active participation of the strange quark with its mass of order the QCD scale, or some other   dynamics peculiar to hadronic
weak interactions, is unknown. Our hope is that through experimental studies of the $\Delta S=0$ HWI with PV observables and through their
theoretical interpretation within the EFT framework that makes for the closest possible contact with QCD, we will gain new insights into the
low-energy weak interactions of the three lightest quarks. In this review, we hope to have sketched a useful roadmap for future progress in this
direction.

As we have emphasized throughout this article, the forefront in this endeavor lies in the arena of single-nucleon and few-body nuclear systems.
Experimental developments have paved the way for completion of precise measurements of the ${\cal O}(10^{-7})$ PV effects in such systems, and
plans are underway for several new few-body experiments. The immediate theoretical challenge is to compute the few-body PV observables using the
EFT framework, allowing one to extract robust values for the low energy constants from experiment. In the longer term, we would like to
determine the extent to which these results are consistent with Standard Model-based expectations, both from the standpoint of symmetries and
from first principles lattice calculations.  In light of  the substantial experimental and theoretical progress in the field  as well as the
significant new opportunities made possible by this progress, we are optimistic that this effort can be successful.

\newpage
\section{Acknowledgements}

The authors would like to thank J. Carlson, B. Desplanques, M.~T. Gericke, G.~Gwinner, G.~L.~ Greene, W.C. Haxton, B.~R.~ Holstein, C.-P. Liu,
C. Maekawa, W.~M.~Snow, U. van Kolck and S.-L. Zhu,  for helpful discussions. This work was supported in part under DOE contract FG02-05ER41361,
NSF Award PHY-0071856, and by grants from NSERC (Canada)  and the NSF.

\bibliography{parity}
\bibliographystyle{annrev}

\end{document}